\documentclass[preprint,11pt]{JHEP3}

\JHEPspecialurl{http://jhep.sissa.it/JOURNAL/JHEP3.tar.gz}

\usepackage{epsfig,multicol,amsmath}

\usepackage{epstopdf}
\usepackage{bm}  
\usepackage{amsmath}     

\usepackage{graphicx}
\usepackage{rotating}
\usepackage{cite}

\def\beq{\begin{equation}}
\def\eeq{\end{equation}}
\def\bea{\begin{eqnarray}}
\def\eea{\end{eqnarray}}

\def\eq#1{{Eq.~(\ref{#1})}}
\def\fig#1{{Fig.~\ref{#1}}}
\newcommand{\bas}{\bar{\alpha}_S}
\newcommand{\as}{\alpha_S}

\newcommand{\Lb}{\left(}
\newcommand{\Rb}{\right)}
\newcommand{\h}{\frac{1}{2}}
\parskip=\itemsep               

\newcommand{\nn}{\nonumber}

\newcommand{\ga}{\gamma}

\newcommand{\De}{\Delta}

\newcommand{\la}{\lambda}

\newcommand{\om}{\omega}

\newcommand{\f}{\frac}
\newcommand{\lab}{\label}

\def\pom{{I\!\!P}}

\title{High density QCD and nucleus-nucleus scattering deeply in the saturation region }
\author{\Large Andrey Kormilitzin${}^{a}$ \thanks{Email: andreyk1@post.tau.ac.il.},\,\,
Eugene\, Levin${}^{a, b}$ \thanks{Email: leving@post.tau.ac.il., eugeny.levin@usm.cl}\,\,\,and \,\, Jeremy\,S.\,Miller${}^{a, c}$
\thanks{Email:  jeremy.miller@ist.utl.pt., jeremymi@post.tau.ac.il}
\\
${}^a$ \, Department of Particle Physics, School of Physics and Astronomy,
Tel Aviv University, Tel Aviv, 69978, Israel\\
${}^b$\, Departamento de F\'\i sica, Universidad T\'ecnica
Federico Santa Mar\'\i a, Avda. Espa\~na 1680,
Casilla 110-V,  Valparaiso, Chile\\
${}^c$\,CENTRA, Departamento de F\'\i sica, Instituto Superior T\'ecnico (IST), Av. Rovisco Pais, 1049-001 Lisboa, Portugal
}

\abstract{ In this paper we solve the equations that describe nucleus-nucleus scattering,
in high density QCD, in the framework of the BFKL Pomeron Calculus.
We found that (i) the contribution of  short distances to the opacity for nucleus-nucleus scattering dies at high energies, (ii) the opacity tends to unity at high energy, and (iii) the main
contribution that survives comes from soft (long distance) processes for large values of the impact parameter.
The corrections to the opacity $\Omega\Lb Y,b\Rb = 1$ were calculated and it turns out that they have a completely different form, namely
( $1  - \Omega \to \exp\Lb - Const\,\sqrt{Y}\Rb$)  than the opacity that stems from the Balitsky-Kovchegov equation, which is
( $1  - \Omega \to \exp\Lb - Const\,Y^2\Rb$).  We reproduce the formula for the nucleus-nucleus cross section
 that is commonly used in the description of nucleus-nucleus scattering, and there is
 no reason why it should be correct in the Glauber-Gribov approach.}

\keywords{high density QCD, BFKL Pomeron, non-linear equation, nucleus-nucleus collisions}
\dedicated{PACS: 13.85.-t, 13.85.Hd, 11.55.-m, 11.55.Bq}
\preprint{TAUP  2919/10\\
{\tt }\\
\today}

\begin{document}

\section{Introduction}
High density QCD  has been developed for the scattering of the dilute system of partons with the dense system of partons \cite{GLR,MUQI,MV,MUCD,B,K,JIMWLK,MBK}.
The typical example of such processes are deep inelastic scattering at low $x$ in which one colorless dipole scatters with the dense target, and/or the hadron-nucleus
interaction where the dilute system of partons in a hadron interacts with the dense partonic system of the target. The main non-linear equations that govern these
interactions have been found \cite{B,K,JIMWLK} and studied in great detail. On the other hand, the scattering of the dense system of partons with the dense system of
 partons has been actively studied \cite{BRA,KOLU,LELU,BIIT,HIMST,MMSW,LMP,ELRE} but with limited success, in spite of the fact that this scattering is closely
related to nucleus-nucleus scattering. The latter is a well known  process that has been studied at RHIC experimentally, and in which the key property of high
density QCD has been observed.

In this paper we revisit the nucleus-nucleus interaction in the framework of high density QCD.
We assume that the dense-dense system interacts at high energy with the effective Lagrangian that can be obtained from
 the BFKL Pomeron Calculus. The final form of the effective action for this type of Pomeron interaction was formulated in Ref. \cite{BRA}
which we will use in this paper. We will discuss the BFKL Pomeron Calculus \cite{BFKL,LI} in the next section, and we will show
 that this approach provides the
set of equations that describes nucleus-nucleus interactions, which was originally derived in Ref.\cite{BRA} (see also  Ref.\cite{KLP} and
 \footnote{Unfortunately, the set of equations derived in Ref.\cite{BRA} has not attracted the attention that it deserves, taking into account that
it is the first theoretical example of treating the dense-dense system of scattering. We can mention only two papers where the numerical solutions to the equations
  have been discussed and some properties of the general solution have been suggested (see Refs.\cite{BOMO,BOBR}). Part of these properties are based on the solution
 for the BFKL Pomeron Calculus in zero transverse dimensions, which does not reflect the analytical solution for this problem given in Ref.\cite{GKLM}} ).

We solve the main equations for the  nucleus-nucleus interaction, which is possible thanks to
the  success in solving the nucleus-nucleus interaction in the  framework of the
BFKL Pomeron Calculus in zero transverse dimensions.
This Calculus models the real QCD case, if we neglect the fact that the sizes of the interacting dipoles could
 change during the ``one-dipole'' to ``two-dipoles'' decay \cite{MUCD}.
The equations that have been proven in the kinematic region \cite{SCHW,GKLM} are:
\beq \label{KR10}
g\,S_{A}(b)\,G_{3\pom}\,e^{\omega(0) Y} \,\,\propto\,\,g\,G_{3\pom} A^{1/3}_i\,e^{\omega(0) Y} \,\,\approx \,\,1;\,\,\,\,\,\,\,\,
G^2_{3\pom}\,e^{\omega(0) Y} \,\,\ll\,\,1
\eeq
where $g_i$ is the vertex of the Pomeron-nucleon interaction, $G_{3 \pom}$ is the triple Pomeron vertex, $S_A(b)$ is the number density
 of nucleons at fixed impact parameter $b$  and  $Y = \ln(1/x)$ is the rapidity.
In this region the nucleus-nucleus scattering amplitude reduces to the sum over the class of net diagrams shown in \fig{netset}.

\FIGURE[ht]{
\centerline{\epsfig{file=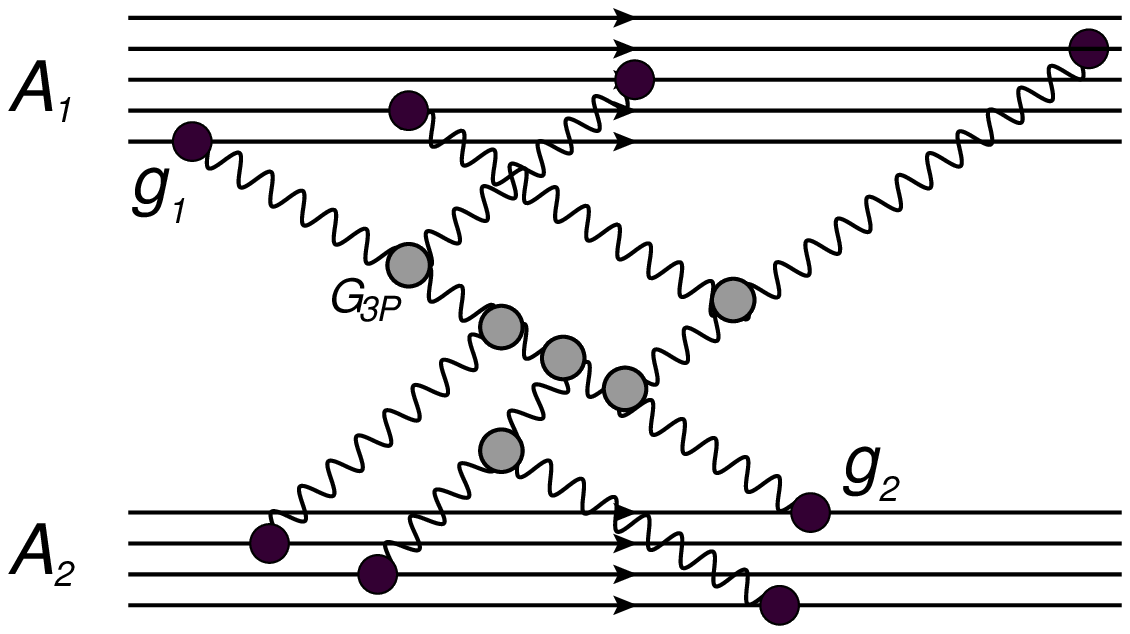,width=90mm}}
\caption{The full set of the diagrams that contribute to the scattering amplitude in the kinematic region \protect\eq{KR10} or \protect\eq{KR1}.}
\label{netset}}


\section{The BFKL Pomeron Calculus}

\subsection{Main ingredients of  the BFKL Pomeron Calculus}
The main ingredients of the BFKL Pomeron Calculus are the same as in the Pomeron Calculus in zero transverse dimensions
(see \fig{netset}), including the Green function of the Pomeron and the vertices of the Pomeron interaction.
In the BFKL Pomeron Calculus only the triple Pomeron vertex and the vertex of the Pomeron interaction with the nucleon contribute.
The first one can be calculated in perturbative QCD which is discussed later on,  while the Pomeron - nucleon vertex is a pure
non-perturbative input. However, in order to make all of the calculations more transparent, we will assume the model where the nucleus is
 a bag of $A$ onia,
where each of them consists of a heavy quark and antiquark. The typical distances in the onium are small and of the order of $1/m_Q$, where $m_Q$ is the heavy quark mass. Therefore, in this oversimplified model we can perform all estimates in the framework of perturbative QCD.

\DOUBLEFIGURE[ht]{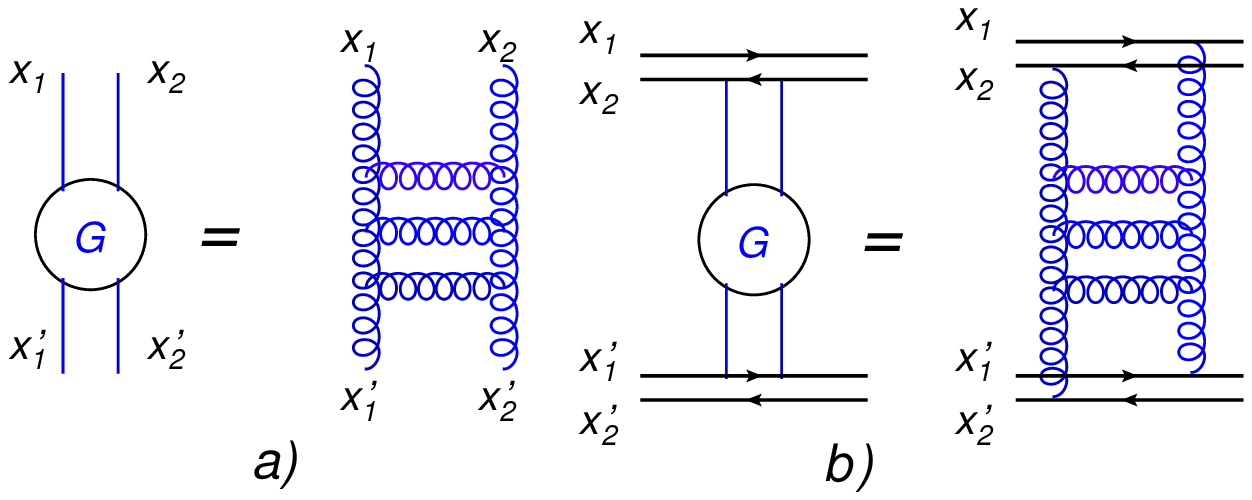,width=70mm,height=35mm}{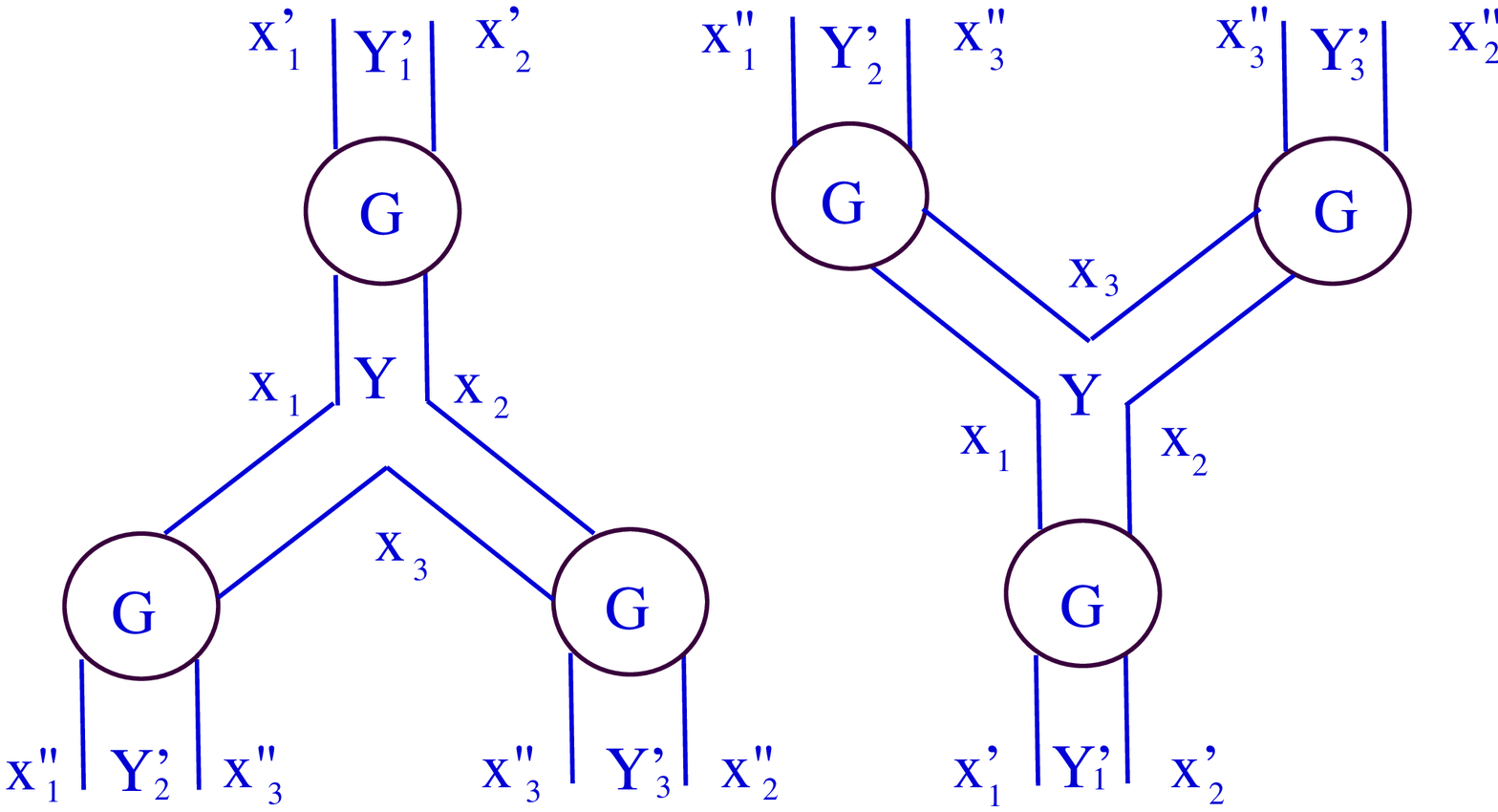,width=60mm,height=35mm}{
The graphical form of the BFKL Pomeron Green function in the coordinate representation (\protect\fig{pomcal1}(a))
and the amplitude for onium-onium scattering (\protect\fig{pomcal1}(b)).\label{pomcal1}}{The graphical form of the triple Pomeron vertex in the coordinate representation.\label{pomcal2}}


One of the main ingredients of this approach is the Green function of the BFKL Pomeron $G\Lb x_1,x_2|x'_1,x'_2\Rb$ (see \fig{pomcal1}).
The explicit expression for this Green function can be found in Ref.\cite{LI}.
The onium-onium forward scattering amplitude can be easily written using the Green function in the following form;

\beq \label{OA}
A\Lb Y = ln s,t=0 \Rb\,\,=\,\,\as^2\int d^2 b \int d^2 r\,\int d^2 R \, |\Psi\Lb r\Rb|^2\, |\Psi\Lb r\Rb|^2\,G\Lb x'_1,x'_2;0|x_1,x_2;
Y\Rb
\eeq
where $\as$ is the QCD coupling , $\Psi$ is the wave function of the onium, $b$ is the impact parameter which
 is equal to $\vec{b}\,=\,\frac{1}{2}\Lb \vec{x}_1 + \vec{x}_2 + \vec{x}'_1 + \vec{x}'_2\Rb$ and $\vec{r} = \vec{x}_1 - \vec{x}_2$
 and $\vec{R} = \vec{x}'_1 - \vec{x'}_2$ are the transverse sizes of the two scattering colorless dipoles.
The Green function has a simple form in this particular case, namely
\beq \label{GFF}
G\Lb r, R ; t=0,Y\Rb\,\,\equiv\,\,\int d^2 b\,G\Lb x_1,x_2;0|x_1,x_2;Y\Rb \,\,=\,\,\int^{i\epsilon + \infty}_{i \epsilon - \infty} \frac{d \nu}{ 2 \pi}\,\lambda\Lb 0,\nu\Rb\,\frac{1}{\sqrt{ r^2\,R^2}}\,\Lb \frac{r^2}{R^2}\Rb^{  i \nu}\,e^{\omega(\nu)\,Y}
\eeq
The energy levels $\omega(\nu)$  are the BFKL eigenvalues
\begin{equation} \label{BFKLOM}
\omega(\nu)\,=\,2\,\bar{\alpha}_S \left( \psi(1) - Re{\, \psi\left(\frac{1}{2} + i
\nu\right)}\right)\,\,=\,\,\bas \left( 2 \psi(1) - \psi\Lb \gamma\Rb  - \psi\Lb 1 - \gamma\Rb \right)\,\,=\,\,\bas\,\chi\Lb \gamma\Rb
\end{equation}
where $\psi(z) = d \ln \Gamma(z)/d z$ and $\Gamma(z)$ is the Euler gamma function,   $\gamma = \h + i \nu$
is the anomalous dimension,  and $\bas= \as N_c/\pi$ where $N_c$ is the number of colors. Finally
\begin{equation} \label{BFKLLA}
\lambda(0,\nu)\,=\frac{1}{[ 1 + 4 \nu^2]^2}
\end{equation}
We do not need to know the explicit expression for the more general case. Instead of this we will discuss some key properties
of the more generalized Green function.  However before this
we give the expression for the triple Pomeron interaction \cite{BRA}
 which can be written in
the coordinate representation for the following process.
Two reggeized gluons with coordinates $x'_1$ and $x'_2$ at rapidity $Y'_1$, which we denote as
$\left\{ x^{\,\prime}_1,x^{\,\prime}_2\vert Y^{\,\prime}_1\right\}$,
decays into two gluon pairs  $\left\{ x^{\prime \prime}_1,x^{\prime \prime}_3 \vert Y'_2\right\}$ and
$\left\{ x^{\prime
\prime}_2,x^{\prime \prime}_3\vert Y'_3\right\}$ due to the
Pomeron splitting at rapidity $Y$. This Pomeron splitting is shown pictorially in \fig{pomcal2}.

\bea
&&2\frac{\pi\,\bar{\alpha}^2_S}{N_c}\,\int\,\frac{d^2 x_1\,d^2\,x_2
\,d^2
x_3}{x^2_{12}\,x^2_{23}\,x^2_{13}}\,\left(L_{12}\,G(x'_1,x'_2;Y'_1|x_1,x_2;
Y)  \right) \times \label{BFKL3P}\\
&&\times G(x_1,x_3;Y|x^{\prime\prime}_1,
x^{\prime\prime}_3;Y'_2)\,G(x_3,x_2;Y|x^{\prime\prime}_3,x^{\prime\prime}_2;Y'_3)\nn\\
\nn\\\mbox{where}\hspace{1cm}
&& L_{12}\,\,\,=\,\,x^4_{12} \,p^2_1\,p^2_2\,\,\,\hspace{0.5cm}\mbox{with}\hspace{0.5cm}
\,\, p^2 \,=\, - \nabla^2 \,\,\,\hspace{0.5cm}\mbox{and}\hspace{0.5cm}\,\, \,x^2_{12}\,\,=\,\,\Lb \vec{x}_1 \,-\,\vec{x}_2\Rb^2\hspace{2cm}\label{L12}\eea

For further presentation we require some properties of the BFKL Green function, which are listed below
\cite{LI}:

\begin{enumerate}
\item
The general definition of the Green function leads to

\bea
&&G^{-1}(x_1,x_2;Y| x'_1,x'_2;Y')\,\, \,=\,p^2_1\,p^2_2\,\left(
\frac{\partial}{\partial Y} + H \right) \,\,=\,\,\left(
\frac{\partial}{\partial Y} + H^+ \right)\,p^2_1\,p^2_2 \label{G1}
\\
\nn\\
&&
H f(x_1,x_2;Y) \,\,= \,\,\frac{\bas}{2
\pi}\,\int\,d^2 x_3\, K\Lb x_1, x_2|  x_3\Rb\,\left\{
f(x_1,x_2;Y)\,-\,f(x_1,x_3;Y)\,-\,f(x_3,x_2;Y) \right\} \hspace{1cm} \label{H}\\
\nn\\
&& K\Lb x_1, x_2|  x_3\Rb\,\,=\,\,\frac{
x^2_{12}}{x^2_{23}\,x^2_{13}}
\label{K}\eea

\item The initial Green function $G_0$ that corresponds to the exchange of two gluons, is
 equal to
\begin{equation} \label{G0}
G_0(x_1,x_2;Y| x'_1,x'_2;Y)\,\,=\,\, \pi^2\,\ln\Lb
\frac{x^2_{1,1'}\,x^2_{2,2'}}{x^2_{1,2'}\,x^2_{1',2}}\Rb  \,\ln
\Lb\frac{x^2_{1,1'}\,x^2_{2,2'}}{x^2_{1,2}\,x^2_{1',2'}}\Rb
\end{equation}
 This form of $G_0$ has been discussed in Ref.\cite{LI}.

\item It should be stressed that
\bea \label{G01}
&&\nabla^2_1 \,\nabla^2_2 \,G_0(x_1,x_2;Y| x'_1,x'_2;Y)=\\
&&=(2\,\pi)^4\,\,\left(\delta^{(2)}( x_1 - x'_1)\,\delta^{(2)}( x_2
- x'_2)\,+\,\delta^{(2)}( x_1 - x'_2)\,\delta^{(2)}( x_2 -
x'_1)\,\right)\nn
\eea

\item At high energy the Green function is also the eigenfunction of the operator $L_{12}$:
\beq \label{LG} L_{12}\,G(x_1,x_2;Y|
x'_1,x'_2;Y')\,\,=\,\,\frac{1}{\lambda(0,\nu)} \,G(x_1,x_2;Y|
x'_1,x'_2;Y')\,\,\approx\,\,G(x_1,x_2;Y| x'_1,x'_2;Y') ;
\eeq

\end{enumerate}

 The last equation holds only approximately in the region where
$\nu \,\ll\,1$, but this is the most interesting region which is
responsible for the high energy asymptotic behavior of the scattering
amplitude.

\subsection{Functional integral formulation of the BFKL Pomeron Calculus.}


The theory with the interaction given by Eq.~(\ref{GFF}) and  Eq.~(\ref{BFKL3P})
can be written through the functional integral \cite{BRA}
\begin{equation} \label{BFKLFI}
Z[\Phi, \Phi^+]\,\,=\,\,\int \,\,D \Phi\,D\Phi^+\,e^S \,\,\,\hspace{0.5cm}\mbox{with}\hspace{0.5cm}\,S \,=\,S_0
\,+\,S_I\,+\,S_E
\end{equation}
where $S_0$ describes free Pomerons, $S_I$ corresponds to their mutual interaction
while $S_E$ relates to the interaction with the external sources (target and
projectile). From Eq.~(\ref{BFKL3P}) and  Eq.~(\ref{BFKLFI}) it is clear that
\begin{equation} \label{S0}
S_0\,=\,\int\,d Y \,d Y'\,d^2 x_1\, d^2 x_2\,d^2 x'_1\, d^2 x'_2\,
\Phi^+(x_1,x_2;Y)\,
G^{-1}(x_1,x_2;Y|x'_1,x'_2;Y')\,\Phi(x'_1,x'_2;Y')
\end{equation}
\beq \label{SI}
S_I\,=\,\frac{2\,\pi \bas^2}{N_c}\,\int \,d Y'\,\int
\,\frac{d^2 x_1\,d^2 x_2\,d^2 x_3}{x^2_{12}\,x^2_{23}\,x^2_{13}}\,
\{ \left( L_{12}\Phi(x_1,x_2;Y')\,\right)\,\Phi^+(x_1,x_3;Y')\,\Phi^+(x_3,x_2;Y')\,\,+\,\,h.c. \} \eeq
For $S_E$ we have the local interaction both in terms of rapidity and coordinates, namely

\begin{equation} \label{SE}
S_E\,=\,-\,\int \,dY'\,d^2 x_1\,d^2 x_2\, \{
\Phi(x_1,x_2;Y')\,\tau_{pr}(x_1,x_2;Y)\,\,+\,\,\Phi^+(x_1,x_2;Y')\,\tau_{tar}(x_1,x_2;Y)
\}
\end{equation}
where $\tau_{pr}$ ($\tau_{tar}$)  stands for the projectile and target, respectively.
The form of the functions $\tau$  depend on the non-perturbative input in our problem.
In our simple model of the nucleus they have the following form

\beq \label{TAU}
\tau_{pr}\,=\,\, \delta\Lb Y - Y'\Rb \,|\Psi\Lb x_{12}\Rb|^2 \,S_{A_1}\Lb \vec{B} - \vec{b}\Rb;\,\,\,\,\,
\tau_{tar}\,=\, \,\delta\Lb Y '- 0\Rb \,|\Psi\Lb x_{12}\Rb|^2 \,S_{A_2}\Lb \vec{b}\Rb
\eeq
where $Y=ln s$ and  $S_{A_i}\Lb b\Rb$ is the number of nucleons at a given impact parameter $b$ in the nucleus $A_i$.
$\vec{B}$ is the impact parameter between the centers of the two nuclei and $b$ is the position of the interacting nucleon (onium)
in the target.
In \eq{TAU} we neglect the impact parameter of the onium - onium interaction in comparison with the impact parameters of the nucleons (onia)
in the nuclei.  Indeed,  the impact parameter of the onium - onium interaction is of the order of the typical onium (nucleon) size,
which is much less than the nucleus size. This is a key assumption of the Glauber approach \cite{GLAUB} which restricts the value of energy until
we can trust this approach. The radius of the onium-onium interaction increases with energy, and at ultra high energy it will be larger than the nucleus size.
A discussion of this increase in the framework of QCD can be found in Ref. \cite{KOWI}. This assumption means
that we can integrate over all impact parameters in the Pomeron interaction, and actually all fields
 $\Phi\Lb x_1,x_2\Rb$ and $\Phi^+\Lb x_1,x_2\Rb$ can be considered to be the fields that depend only on the dipole sizes
 ($\Phi\Lb x_{12}\Rb$ and
$\Phi^+\Lb x_{12}\Rb$).

\subsection{The equation for nucleus-nucleus scattering}

\fig{glaub}  shows the Glauber re-scatterings that give the largest contribution to the nucleus-nucleus scattering amplitude.
The onium (nucleon) - onium (nucleon) scattering amplitude is of the order of $\as^2$ and the exchange of one BFKL Pomeron leads
 to the following contribution to the nucleus-nucleus scattering amplitude

\bea \label{GLA10}
A_{\pom}\Lb Y,B\Rb  &=&  \as^2 \int d^2 b\, S_{A_1}\Lb \vec{B} - \vec{b}\Rb\,S_{A_2}\Lb b \Rb\,\int d^2 r\, d^2 R \,
|\Psi\Lb r\Rb|^2\,
 |\Psi\Lb R\Rb|^2\,\int d^2 b'\,G\Lb R;0|r, b';Y\Rb\hspace{1cm}\\
 &=&  \as^2 \!\int d^2 b\, S_{A_1}\Lb \vec{B} - \vec{b}\Rb\,S_{A_2}\Lb b \Rb\,\int d^2 r\,|\Psi\Lb r\Rb|^2\,{\cal G} \Lb r ; Y\Rb\,\propto   \,\as^2\,A^{1/3}_1\,A^{1/3}_2\,A^{2/3}_1 \mbox{(for $A_1 <  A_2$)}\,e^{\omega(0)\,Y}\nn
\eea
Notice that the factor $A^{2/3}_1 $ stems from the  $b$ integration. The contribution of the triple 
Pomeron interaction leads to the amplitude shown in \fig{g13p}. Using    \eq{BFKL3P} and the function ${\cal G} \Lb r ; Y\Rb$
 that has been introduced in  \eq{GLA10}, this contribution
can be rewritten in the form
\bea
&& 2\bas \as^4  \int \!d^2 b\, S_{A_1}\Lb \vec{B} - \vec{b}\Rb\,S^2_{A_2}\Lb b \Rb \int^Y_0 \!\!d Y'\int\,\frac{d^2 x_1\,d^2\,x_2
\,d^2
x_3}{x^2_{12}\,x^2_{23}\,x^2_{13}}\,L_{12}\,{\cal G}\Lb x_{12}; Y - Y' \Rb\,\label{GLA101}\\
&& \times {\cal G}\Lb x_{23}; Y'  - 0\Rb\,{\cal G}\Lb x_{23}; Y'  - 0\Rb
\, \,\,\propto\,\,\,\,\Big( \as^2\,A^{1/3}_1\,A^{1/3}_2\,A^{2/3}_1 \,e^{\omega(0)\,Y}\Big)\,\Big( \as^2 A^{1/3}_2\,e^{\omega(0)\,Y}\Big)\nn\\
&&\,\,\sim \,\,\,\,\,A_{\pom}\Lb Y,B\Rb\Big( \as^2 A^{1/3}_2\,e^{\omega(0)\,Y}\Big)\nn
\eea

\eq{GLA101} shows that if
\beq \label{KR1}
\as^2\,A^{1/3}_1\,A^{1/3}_2\,e^{\omega(0)\,Y}\,\,\sim\,\,1; \,\,\,\mbox{but}\,\,\,\,\as^2\,A^{1/3}_1\,e^{\omega(
0)\,Y}\,\,\leq\,\,1
\eeq
the diagrams with triple Pomeron interactions are small and only
the Glauber - type re-scatterings shown in \fig{glaub}, give the largest contribution to the nucleus-nucleus scattering amplitude. However, if
\beq \label{KR2}
\as^2\,A^{1/3}_1\,e^{\omega(0)\,Y}\,\, \sim\,\,1\,\,\,\mbox{while} \,\,\,\as^2\,e^{\omega(0)\,Y}\,\,\leq\,\,1
\eeq
the ``net'' diagrams of \fig{netset} will also contribute. The set of ``net'' diagrams (see \fig{netset}) has two topological
 characteristics: (1) these diagrams are two nuclei irreducible or, in other words, any diagram cannot be redrawn
  as   two parts separated by two nuclei states in  the $s$-channel; and (2) the smallness of each triple Pomeron vertex  
($G_{3\pom} \propto \as^2$)
is compensated by a large factor of $A^{1/3}_1$ or $A^{1/3}_2$.  The smallness of $\as^2\,e^{\omega(0)\,Y}$  leads to the fact that we can neglect the BFKL Pomeron loops or corrections to Pomeron vertices which are of the order of
$\int dY_1 d Y_2 G_{3\pom}(Y_1)\,G_{3\pom}(Y_2)\exp\Lb \omega(0) (Y_1 - Y_2)\Rb \propto \as^2 \exp\Lb \omega(0) Y\Rb$.

It is obvious that in order to sum both the Glauber re-scatterings and the net
diagrams, we need
to search for the nucleus-nucleus scattering amplitude in the form\footnote{The proof of this form can be found in Ref.\cite{BRA}
in the framework of the BFKL Pomeron Calculus and in Ref. \cite{GKLM}, where an explanation can be found 
of how \eq{GLA1} derives from the parton cascade in the Pomeron Calculus in zero transverse dimensions.}

\bea
&&A\Lb AA; s, B\Rb\,\,=\,\,i \Big(1\,\,-\,\,\exp\Lb - \h  \Omega\Lb s,B\Rb\Rb\Big)\label{GLA1}\\
\nn\\
 \mbox{where}\hspace{2cm}
&&\Omega\Lb s,B\Rb\,\,=\,\,A_{\pom}\Lb Y,B\Rb\,\,+\,\,\dots \,=\, \mbox{net diagrams}\label{GLA2}\\
&&=\as^2 A^{1/3}_1\,A^{1/3}_2 A^{2/3}_1 s^{\omega(0)} \Sigma\Big( \as^2 A^{1/3}_1\,s^{\omega(0)} , \as A^{1/3}_2 e^{\omega(0)}\Big) \hspace{1cm} \mbox{(for $A_1 < A_2$)}\,\nn
\eea

Indeed, the Glauber form of \eq{GLA1} sums all two nuclei reducible diagrams (see \fig{glaub} and \fig{g13p}) while net diagrams,
being two nuclei irreducible, provide the contribution $\Omega/2$ (see \fig{g13p}). The normalization of \eq{GLA1} is the following
\beq \label{GLA3}
\sigma_{tot}\Lb AA; s\Rb \,\,=\,\,2 \,\int d^2 B \,\mbox{Im} A\Lb AA; s, B\Rb
\eeq

\DOUBLEFIGURE[ht]{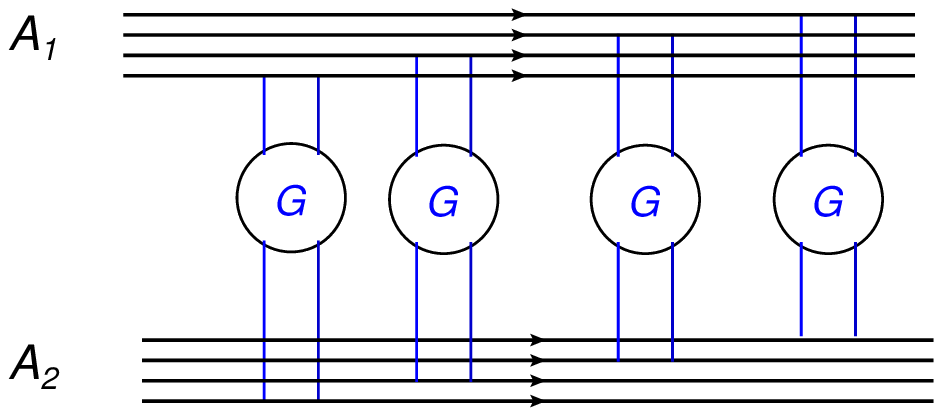,width=70mm,height=30mm}{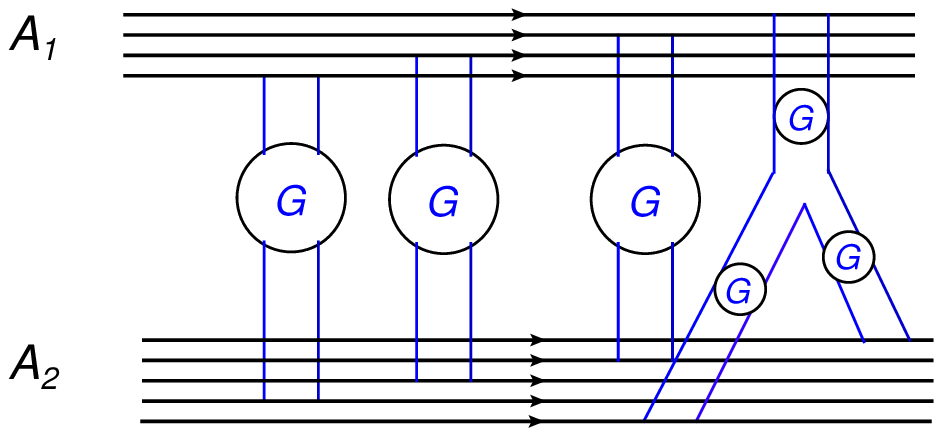,width=70mm,height=30mm}{
The Glauber-type re-scattering for the nucleus-nucleus scattering amplitude\label{glaub}}{The contribution of the triple Pomeron vertex
to the Glauber-type formula.\label{g13p}}


The equations that sum the net diagrams of \fig{netset} can be derived from \cite{BRA,KLP}  from the following equations of motion

\beq \label{EQA1}
\frac{\delta S}{\delta \Phi\Lb x,y\Rb}\,\,=\,\,0;\,\,\,\,\,\,\,\,\frac{\delta S}{\delta \Phi^+\Lb x,y\Rb}\,\,=\,\,0;
\eeq

Using  the action of \eq{BFKLFI}, one can derive from \eq{EQA1} the system of two equations of motion for the theory given by \eq{BFKLFI}:

\bea
&& \frac{\partial \langle \Phi(x,y;Y')\rangle}{\partial\,Y'}\,\,= \label{EQF21}\\
 &&\frac{\bas}{2
\pi}\!\int\! d^2 z\,K\left(x,y|z
\right) \Big\{ \langle \Phi(x,z;Y') \rangle +\langle \Phi(z,y;Y')\rangle -\langle \Phi(x,y;Y')\rangle -4 \pi \as \langle \Phi(z,y;Y')
\Phi(z,y;Y') \rangle\Big\} \nonumber
\\
&&- \frac{1}{(2 \pi)^4}\frac{\bas}{ \pi} 4 \pi \as \int d^2z \frac{d^2 x' d^2\,y'}
{(x' - y')^4} G_0\left(x,y;Y'|x',y';Y'\right) K \left(x',y'|z
\right) \langle \left\{L_{zy'}  \Phi^+\left(z,y',Y-Y'\right)\right\} \Phi(x',z;Y')\rangle \nonumber\\
\nn\\
\nn\\
 && -\frac{\partial \langle \Phi^+(x,y;Y - Y')}{\partial\,Y'}= \label{EQF22} \\
&&=\frac{\bas}{2 \pi}\,\int\,d^2 z K\left(x,y|z
\right)\Big\{\langle \Phi^+(x,z;Y - Y')\rangle  + \langle \Phi^+(z,y;Y - Y')\rangle -\langle \Phi^+(x,y;Y - Y')\rangle -
\Big. \nonumber \\
&&-\Big.
 4\pi \as \langle \Phi ^+(z,y;Y -
Y')\Phi^+(z,y;Y
- Y')\rangle
\Big\} \nonumber\\
 &&- \frac{1}{(2 \pi)^4}\frac{\bas}{ \pi}4 \pi \as \int d^2 z\frac{d^2 x' d^2y'}{( x' -
y')^4} G_0\left(x,y;Y'|x',y';Y'\right) K \left(x',y'|z \right) \langle \left\{ L_{zy'} \Phi\left(z,y';
Y'\right)\right\} \Phi^+(x',z;Y - Y')\rangle\,\nonumber
\eea

In \eq{EQF21} and \eq{EQF22} the procedure of averaging is defined  as
\beq \label{AVO}
\langle O(x,z;Y') \rangle \,\,\equiv\,\frac{\int D \Phi D \Phi^+\, O(x,z,Y) \,e^{S\left[\Phi, \Phi^+\right]}}{
\int D \Phi D \Phi^+\, \,e^{S\left[\Phi, \Phi^+\right]}|_{S_E = 0}}
\eeq

\eq{EQF21} and \eq{EQF22} are general equations for the theory of interacting BFKL Pomerons which sum all possible diagrams.
For the net diagrams we have the additional property that
$$
\langle \Phi(x,z;Y)\, \Phi(z,y;Y)\rangle = \langle \Phi(x,z;Y)\rangle\,\langle \Phi(z,y;Y)\rangle ;\,\,\,
\langle \Phi^+(x,z;Y)\, \Phi^+(z,y;Y)\rangle  =  \langle \Phi^+(x,z;Y)\rangle\,\langle \Phi^+(z,y;Y)\rangle;
$$
\beq \label{AVTO}
\langle \Phi^+(x,z;Y)\, \Phi(z,y;Y)\rangle \,\,=\,\,\langle \Phi^+(x,z;Y)\rangle\,\langle \Phi(z,y;Y)\rangle
\eeq
Indeed,  drawing  all diagrams in the kinematic region of \eq{KR2} for  $\langle \Phi(x,z;Y)\, \Phi(z,y;Y) \rangle$ one can see
that we have two sets of diagrams, that are disconnected from each other.
Introducing
\bea \label{NN}
N\Lb x, y; Y\Rb\,\,&=&\,\,- 4 \pi \as \frac{\int D \Phi D \Phi^+\, \Phi(x,z,Y) \,e^{S\left[\Phi, \Phi^+\right]}}{\int D \Phi D \Phi^+\, \,e^{S\left[\Phi, \Phi^+\right]}|_{S_E = 0}}\,\,\,\,\,\,\,\,\mbox{and} \nn\\
N^+\Lb x, y; Y\Rb\,\,&=&\,\,- 4 \pi \as \frac{\int D \Phi D \Phi^+\, \Phi^+(x,z,Y) \,e^{S\left[\Phi, \Phi^+\right]}}{\int D \Phi D \Phi^+\, \,e^{S\left[\Phi, \Phi^+\right]}|_{S_E = 0}}
\eea
and using \eq{AVTO}
we rewrite \eq{EQF21} and \eq{EQF22}  in the form

\bea
 \,\,\,\,\,& & \frac{\partial N(x,y;Y')}{\partial\,Y'}\,\,= \label{EQ21}\\
 &=&\,\,\frac{\bas}{2
\pi}\,\int\,d^2\,z\,K\left(x,y|z
\right)\,\Big\{ N(x,z;Y') \,+\,N(z,y;Y')\,-\,N(x,y;Y')\,- N(z,y;Y')\,N(z,y;Y') \Big\} \nonumber
\\
&-& \,\frac{1}{(2 \pi)^4}\,\frac{\bas}{ \pi}\,\int d^2z\,\frac{d^2 x'\,d^2\,y'}
{(x' - y')^4}\,G_0\left(x,y;Y'|x',y';Y'\right)\,\,K \left(x',y'|z
\right)\,\,\left\{L_{zy'} \,N^+\left(z,y',Y-Y'\right)\right\}\,N(x',z;Y')\,\nonumber\\
\nn\\
\nn\\
 & & -\,\frac{\partial N^+(x,y;Y - Y')}{\partial\,Y'}\,\,=\,\, \label{EQ22} \\
&=&\,\,\frac{\bas}{2 \pi}\,\int\,d^2\,z\,K\left(x,y|z
\right)\,\Big\{N^+(x,z;Y - Y') \,+\,N^+(z,y;Y - Y')\,-\,N^+(x,y;Y - Y')\,- \Big. \nonumber \\
&-&\Big.
\, \, N^+(z,y;Y -
Y')\,N^+(z,y;Y
- Y')
\Big\} \nonumber\\
 &-& \,\,\frac{1}{(2 \pi)^4}\,\frac{\bas}{ \pi}\,\int d^2 z\frac{d^2 x'\,d^2\,y'}{( x' -
y')^4}\,G_0\left(x,y;Y'|x',y';Y'\right)\,\,K \left(x',y'|z \right)\,\left\{ L_{zy'}N\left(z,y';
Y'\right)\right\}\,N^+(x',z;Y - Y')\,\nonumber
\eea

\FIGURE[h]{
\centerline{\epsfig{file=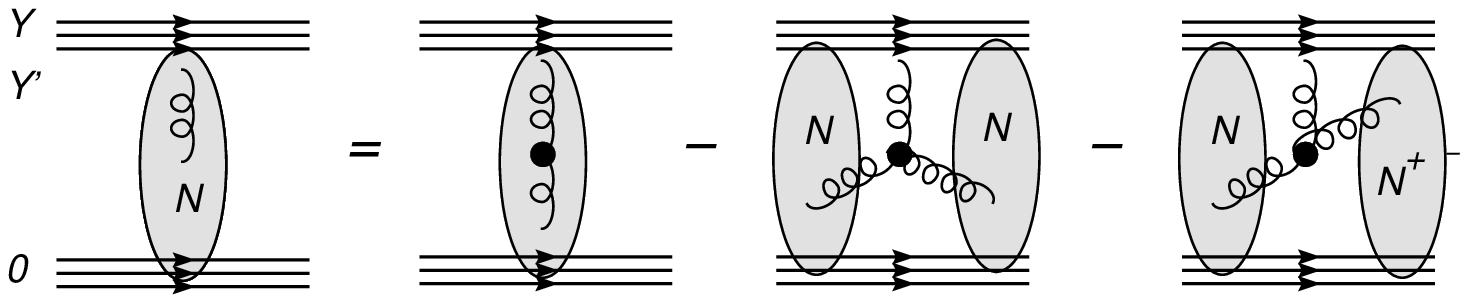,width=150mm}}
\caption{The pictorial representation of  \protect\eq{EQ21} )(see \fig{eqAA}-A). \fig{eqAA}-B shows the initial  condition for this equation.
The black blob denotes the triple gluon vertex while the wavy  lines describe BFKL Pomerons.}
\label{eqAA}
}

\FIGURE[h]{
\centerline{\epsfig{file=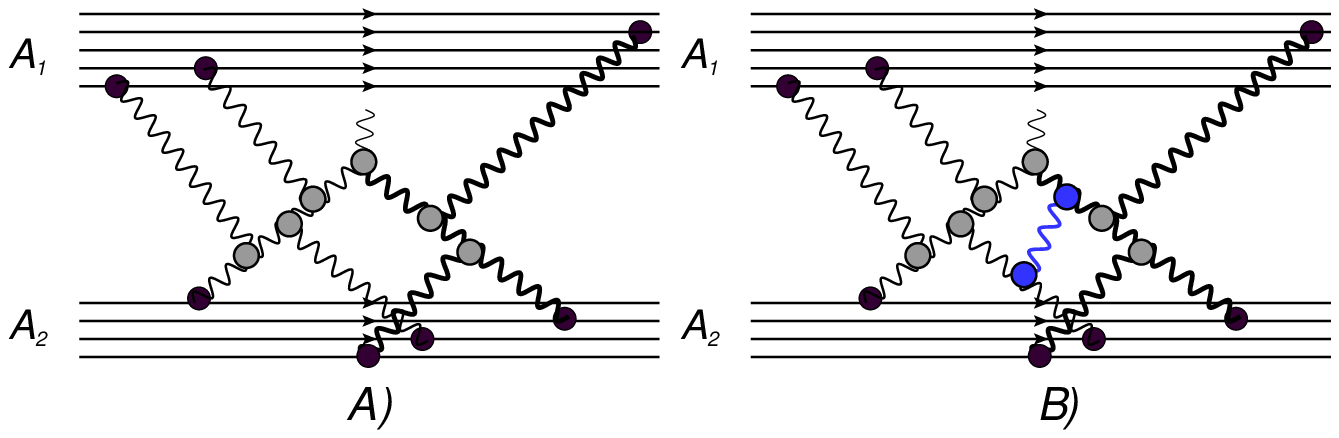,width=150mm,height=45mm}}
\caption{The diagrams for  $N$. In \fig{eqAAil}-A we denote by bold wavy lines and  by normal lines,
 the Pomerons that contribute to the different $N$ and $N^2$ terms of \protect\eq{EQ21} (see \fig{eqAA}-A). In   \fig{eqAAil}-B 
the Pomeron that connects the two sets of  diagrams is shown in blue. One can see that its contribution
 is suppressed by a factor of $\as^2 s^{\omega(0) }$.}
\label{eqAAil}
}
The graphical form of the first equation (\eq{EQ21}) is shown in \fig{eqAA}-A. The  equation for $N^+$ has the same  graphical form
 with the replacement $N \to N^+$ and $N^+ \to N$. Note, that the initial condition for $N^+$ has the form of \fig{eqAA}-B but with the Pomeron attached to the upper nucleus.

Using the initial condition of  \fig{eqAA}-B one can check that the iterations of \eq{EQ21} reproduce the set of net diagrams of
 \fig{netset}. Comparing \fig{netset} with \fig{eqAA} one can see that \eq{EQ21} sums the net diagrams, proving
 that the set of equations of \eq{EQ21} and \eq{EQ22} describe the nucleus-nucleus scattering.  \eq{AVTO} leads
 to the second and the third terms of equation. These terms reflect the key property of the net diagrams, namely that
 three  Pomerons in a triple Pomeron vertex generates the two sets of  diagrams that are
 disconnected from each other (see \fig{eqAAil}-A, where these two sets of diagrams are shown in bold and normal wavy
 lines). In other words, each Pomeron line that connects two  sets of diagrams gives the contribution  
 of the order of $\as^2\,e^{\omega(0)\,Y}\,\,\leq\,\,1$ (see \fig{eqAAil}-B for example).

By assuming that $N^+$ is small, \eq{EQ21} reduces to the familiar Balitsky-Kovchegov (BK) equation.

The typical example of this type of situation is deep inelastic scattering (DIS), but even in this case we have to be careful \cite{MUSH}.
 Indeed,  the BK equation sums the class of ``fan'' diagrams shown in \fig{fandi}.
For large photon virtualities $Q$, it is sufficient to take into account just the scattering of one dipole whose transverse 
size is of the order of $1/Q$. However, the typical sizes of the dipoles inside of the diagrams of \fig{fandi} can reach values, 
such that it is no longer justified to use this class of ``fan'' diagrams \cite{GLR,MUSH}.

It is interesting to notice that in the B-K limit, $N$ being the Green function of two gluons can be viewed as the amplitude of 
the dipole scattering. In the case of nucleus-nucleus scattering the scattering amplitude of a dipole that interacts with two nuclei is a more complicated observable which for small $N$ and $N^+$ is equal to $N + N^+$.

It  should be stressed that the proof of \eq{EQ21} and \eq{EQ22}, as well as the relation to BFKL Pomeron diagrams
 were discussed in Refs.\cite{BRA,KLP,BOMO,BOBR} as well as in Ref.\cite{GKLM} for the BFKL Pomeron Calculus in zero transverse dimensions.
 In this section we just introduce notations and the main ingredients of this approach for completeness of presentation.


\DOUBLEFIGURE[ht]{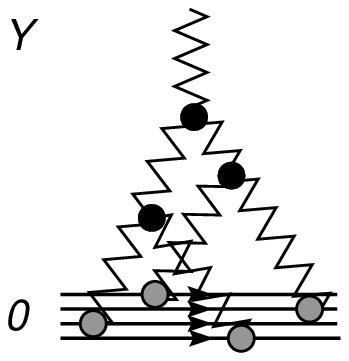,width=40mm}{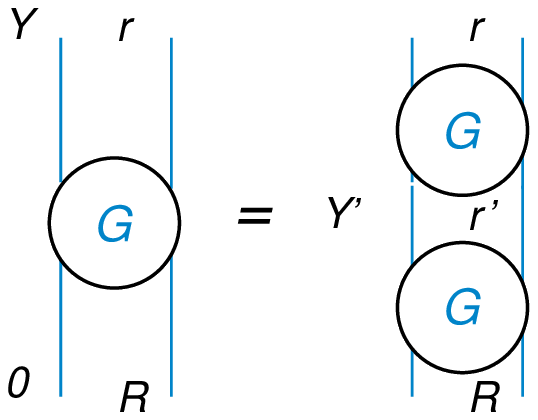,width=60mm,height=40mm}{The ``fan'' diagrams that are summed by the Balitsky-Kovchegov equation. The zig zag lines denote the BFKL Pomeron.\label{fandi}
}{The graphical form of the BFKL contribution of \protect\eq{TS1}.\label{n2}}

\section{Solution to the main equations deeply in the saturation region.}
\subsection{Two saturation scales.}
In Ref.\cite{MUSH}, it is shown that actually we are dealing with two saturation scales even in the dilute-dense scattering system.
Indeed, let us consider the BFKL Pomeron contribution to the case of DIS.
From the complete set of eigenfunctions of the BFKL equation, one can conclude that (see \eq{GFF})
\bea \label{TS1}
&&G\Lb r, R ; t=0,Y\Rb\,\, =\,\,\int\,d^2 r' \,G\Lb r, r' ; t=0,Y - Y'\Rb\,G\Lb r', R ; t=0,Y'-0\Rb\,\\
&&=\,\,\frac{1}{\sqrt{r^2 R^2}}
\int^{i \epsilon + \infty}_{i\epsilon - \infty} \frac{ d \nu'}{2 \pi}\int^{i \epsilon + \infty}_{i\epsilon - \infty} \frac{ d \nu}{2 \pi} \int \frac{d^2 r'}{r'^2}\,\Lb \frac{r^2}{r'^2}\Rb^{i \nu}\,  \Lb \frac{r'^2}{R^2}\Rb^{i \nu'}\exp\Big\{
 \omega(\nu)(Y - Y') \,+\,\omega(\nu')(Y' -0)\Big\}\nn
\eea
Integrating over $r'$ leads to $\delta( \nu - \nu')$ and provides the relation given in \eq{TS1}, which is the general
 property of the Green function. On the other hand, each of the Green functions in \eq{TS1} has its own saturation momentum. It is well known \cite{GLR,QSA,MUTR,MP} that the equation for the saturation scale does not depend on the non-linear terms and can be derived from the knowledge of only the linear part of the equation with the BFKL kernel. Using the general equation for the saturation scale \cite{GLR,QSA,MUTR,MP}, one finds two saturation scales  for the two Green functions
\bea
G\Lb r, r' ; t=0, Y - Y'\Rb\,& \,\,\,\rightarrow \,\,\,&\,\, \ln \Lb r^2/r'^2_s\Rb\,\,=\,\,-\frac{\omega\Lb \gamma_{cr}\Rb}{1 - \gamma_{cr}}\,\Lb Y\,-\,Y'\Rb; \label{TS21}\\
G\Lb r', R ; t=0,  Y'-0\Rb\,& \rightarrow &\,\,\ln \Lb r'^2_s/R^2\Rb\,\,=\,\,-\frac{\omega\Lb \gamma_{cr}\Rb}{1 - \gamma_{cr}}\,\Lb Y\,-\,0\Rb; \label{TS22}
\eea
where $Q^2_s \,=\,1/r^2_s$ and $\gamma = \h + i \nu$ while $\gamma_{cr}$ can be found from the equation
\beq \label{GASAT}
-\,\frac{d \omega\Lb \gamma\Rb}{d \gamma} |_{\gamma = \gamma_{cr}}\,\,=\,\,\frac{\omega\Lb \gamma_{cr}\Rb}{1\,-\,\gamma_{cr}}
\eeq
resolving these equations we obtain two saturation scales:
\bea
\ln\Lb Q^2_{1,s}\,r^2\Rb \,&=&\,-\frac{\omega\Lb \gamma_{cr}\Rb}{1 - \gamma_{cr}}\,\Lb Y\,-\,Y'\Rb  ;\label{TS31}\\
\ln\Lb Q^2_{2,s}\,R^2\Rb \,&=&\,\,\frac{\omega\Lb \gamma_{cr}\Rb}{1 - \gamma_{cr}}\,\Lb Y' - 0 \Rb; \label{TS32}
\eea

\eq{TS31} and \eq{TS32} can be rewritten in the form
\bea
 Q^2_{1,s}\,&=&\,Q^2 \exp\Big(-\frac{\omega\Lb \gamma_{cr}\Rb}{1 - \gamma_{cr}}\,\Lb Y\,-\,Y'\Rb \Big) ;\label{TS41}\\
Q^2_{2,s} \,&=&\,\,Q^2_{2,s}\Lb Y=0\Rb\,\exp\Big( \frac{\omega\Lb \gamma_{cr}\Rb}{1 - \gamma_{cr}}\,\Lb Y' - 0 \Rb \Big);\label{TS42}
\eea

\FIGURE[h]{
\centerline{\epsfig{file=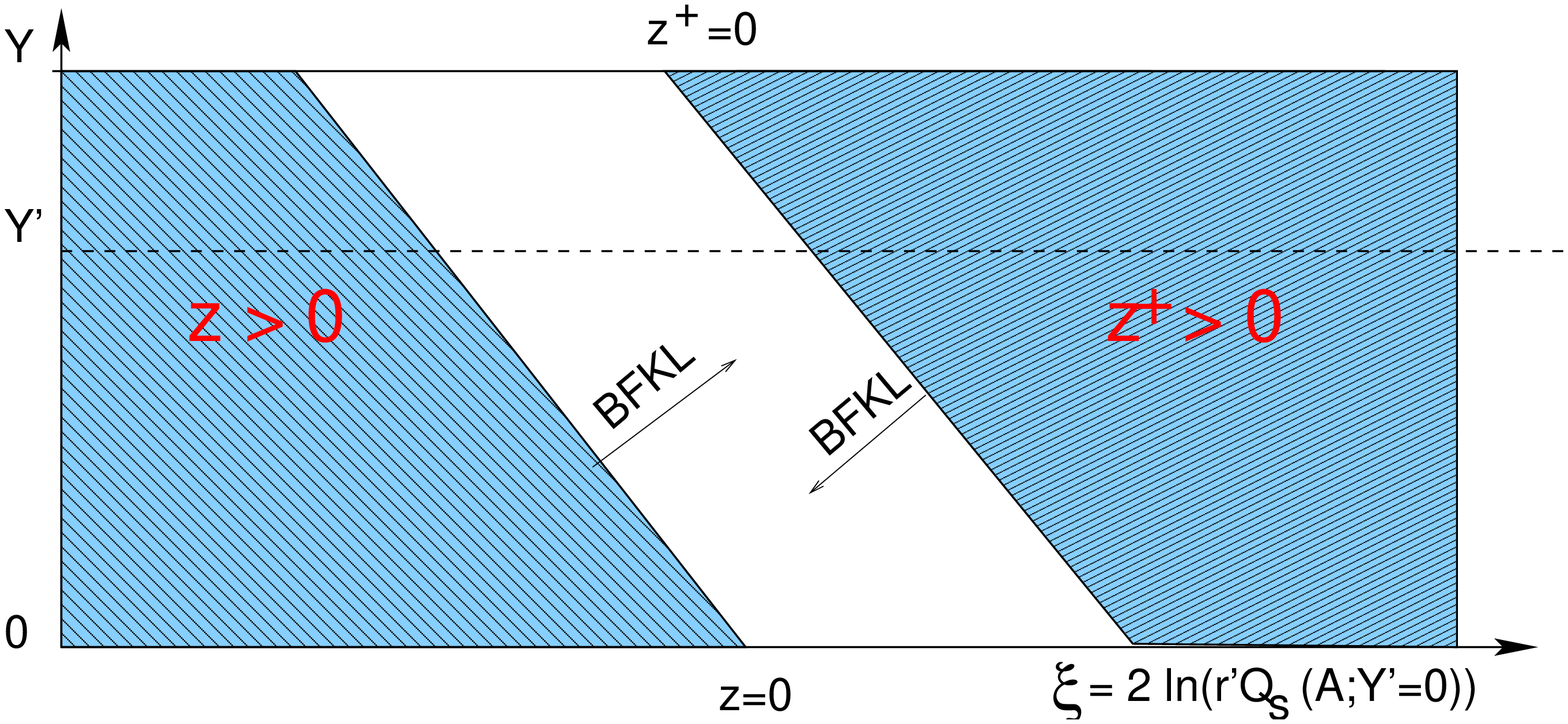,width=120mm}}
\caption{Two saturation scales for DIS.}
\label{ts}
}

It is useful to introduce two scaling variables:

\beq \label{TS5}
z^+\,\,=\,\,\ln\Lb r'^2\,Q^2_{2,s}\Rb \,\,=\,\,\lambda_s\,Y' + \,\xi;\,\,\,\,\,z\,=\,- \ln\Lb r'^2\,Q^2_{1,s}\Rb\,\,=\,\,\lambda_s\,\Lb Y - Y'\Rb - \xi
\eeq
where $\xi \,=\,\ln\Lb r'^2\,Q^2_{2,s} (Y'=0)\Rb$ and $\lambda_s = \omega\Lb \gamma_{cr}\Rb/(1 - \gamma_{cr})$.
The saturation effects start to be essential when both $z$ and $z^+$ are positive.
For negative $z$ and $z^+$ we can safely use the linear (BFKL) evolution equation (see \fig{ts}).
For DIS due to the large value of the photon virtuality, we have the region where the size of the dipole is smaller than both of the inverse saturation scales \cite{MUSH,DG} (see \fig{ts}). However, in the case of nucleus-nucleus scattering the situation is quite different. In this case in the entire kinematic region of low $x$ we cannot neglect the saturation effects, rather we have to solve the non-linear equations (see \fig{tsa}). It is worth mentioning that
\beq \label{TS6}
z\,\,+\,\,z^+\,\,=\,\, \ln\Lb r^2 Q^2_s\Lb Y\Rb\Rb\,\,\equiv\,\,\zeta\,\,\,\,\,\,\,\mbox{where}\,\,\,Q^2_s\,\,=\,\,Q^2_{2,s}\Lb Y = 0\Rb e^{\lambda_s Y}
\eeq
One recognizes that $Q_s$ is the saturation momentum in the Balitsky-Kovchegov equation for DIS. Although the idea of two saturation scales has been on the market during the past six years starting with the paper of
Mueller and Shoshi\cite{MUSH}, it still needs some demystifying. For this purpose let us consider deep inelastic scattering
 at low $x$ in the double log approximation (DLA). The main contribution stems from the kinematic region where
\beq \label{TS7}
  1\,\gg\,x_1\,\gg\,\dots\,\gg\,x_i\,\gg\,\dots\,\gg\, x;\,\,\,\,Q_0\,\ll\,k_{1,\perp}\,\ll\,\dots\,k_{i,\perp}\,\ll\,\dots\,\ll\,k_{n,\perp}\,\ll\,Q;
\eeq

The dipole amplitude in the momentum representation in the DLA  looks as follows
\beq \label{TS8}
N\Lb Q;Q_0; Y\Rb\,\,=\,\,Const \frac{Q^2_0}{Q^2}\,\exp\Big( 2 \sqrt{\bas Y\,\ln\Lb Q^2/Q^2_0\Rb}\Big)
\eeq
and this amplitude satisfies the unitarity constraints $ N < 1$  if
\beq \label{TS9}
Q^2 \,>\,Q^2_s\,\,=\,\,Q^2_0\,e^{4 \bas Y}
\eeq
 \eq{TS1} can be written in the form
\beq \label{TS10}
N\Lb Q;Q_0; Y\Rb\,\,\propto\,\,\int d \ln\Lb k^2_{i,\perp}/Q^2_0\Rb\,N\Lb Q,k_{i,\perp}; Y - Y'\Rb\,N\Lb k_{i,\perp},Q_0;  Y'\Rb
\eeq
The amplitude $N\Lb k_{i,\perp},Q_0;  Y'\Rb \,< \,1$ for
\beq \label{TS11}
k^2_{i,\perp}\,>\,Q^2_s\,=\,Q^2_0\,e^{4 \bas \,Y'}
\eeq
while the amplitude $N\Lb Q,k_{i,\perp}; Y - Y'\Rb\,<\,1$ if
\beq \label{TS12}
Q^2\,>\,k^2_{i,\perp}\,e^{4 \bas(Y - Y')}
\eeq
One can see that the value of $k_{i,\perp}$ for which both amplitudes are in the region where we can apply
 perturbative QCD or, in other words, where both amplitudes are small, has to satisfy the following inequality
\beq \label{TS13}
Q^2_{1,s} \,=\,Q^2\,e^{-\,4 \bas\,(Y - Y')}\,\,>\,\,k^2_{i,\perp}\,\,>\,\,Q^2_{2,s} \,=\,\,Q^2_0\,e^{4 \bas \,Y'}
\eeq

It should be stressed that even when $Q^2$ is large ($Q^2 > Q^2_s$, see \eq{TS9}) and the solution of \eq{TS8}  turns
 out to be small enough to satisfy the unitarity constraints, the partons inside the DGLAP cascade  with transverse momenta given in \eq{TS13},
violate unitarity. If $Q^2 = Q^2_s$ (see \eq{TS9})  both parts of the inequality of \eq{TS13} are equal and all partons violate unitarity. For $Q^2 <Q^2_s$ we have the situation shown in \fig{tsa}. We hope that this equation clarified the appearance of two saturation scales and explains \fig{ts} and \fig{tsa}.\\
 It should be stressed that \eq{TS21} as well as the entire discussion in this section,
 including \fig{ts} and \fig{tsa},  are related  to the partons (dipoles)
 that give the main contribution to the total cross section (BFKL Pomeron).
  To illustrate this point, it is instructive to consider partons with $k_{i, \perp} >> Q$
 and $ k_{i, \perp} >> Q_0$,  for which we expect that saturation effects will be small, which would seem to disagree with \eq{TS13}
 \footnote{We thank our referee, who drew our attention to this point.}.
These partons do not contribute to the deep inelastic scattering in the DLA approach, as we have seen in \eq{TS7}, and therefore,
 we do not consider them.
 However, it is worthwhile to discuss why and how these parton are not essential.
 For these partons that we are discussing, \eq{TS1} looks  like \eq{TS10},  however 
\beq \label{TS14} 
N\Lb Q ; k^2_{i,\perp},Y - Y'\Rb\,\,=\,\,Const \frac{Q^2}{k^2_{i,\perp}}\,\exp\Big( 2 \,\sqrt{\bas (Y\,-\,Y')\,\ln\Lb k^2_{i, \perp}/Q^2\Rb}\,\Big)
\eeq
Taking \eq{TS14} into account, we can rewrite \eq{TS1}  for these partons as
\bea 
&&N\Lb Q;Q_0; Y\Rb\!\!\propto\!\!\int \frac{ Q^2\,d \ln\Lb k^2_{i,\perp}/Q^2_0\Rb}{ k^2_{i,\perp}}\,\exp\Big(
 2 \,\sqrt{\bas (Y\,- \,Y')\,\ln\Lb k^2_{i, \perp}/Q^2\Rb}  + 2\, \sqrt{\bas \,Y'\,\ln\Lb k^2_{i, \perp}/Q^2_0\Rb}\,\Big)\hspace{1cm}
\label{TS15}\eea

Integration over $k_{i,\perp}$ in \eq{TS15}
leads to  $k_{i,\perp} \to Q$, and hence the resulting contribution to 
$N\Lb Q;Q_0; Y\Rb$ is proportional to 
$\exp\Big( 2 \sqrt{\bas Y'\,\ln\Lb Q^2/Q^2_0\Rb}\Big) \,\ll\,\exp\Big( 2 \sqrt{\bas Y\,\ln\Lb Q^2/Q^2_0\Rb}\Big) (\mbox{\eq{TS8}})$.
  In the general case of \eq{TS1}, the integration measure changes from $d^2 r'/r'^2$, to $d^2 r'/\sqrt{r^2\,R^2}$.
Consequently the integration over $r^{\,\prime}$  leads to  $\delta \Lb \nu - \nu'\,-\,i \Rb$ instead of  $\delta \Lb \nu - \nu'\Rb$. 
Thus the $\nu'$ integration leads to the contribution of such partons that forms a  negligible part of the total cross section.
 It
 should be mentioned that a  much more detailed analysis can be found in Ref.\cite{MUSH}. The last item that
 we wish to mention, is that we are discussing the total cross section, whereas the contribution to the inclusive cross section
  has a completely different
form, with two saturation momenta, which are different.

\subsection{Solution deeply in the saturation region : linearized equations}

One can see that in the kinematic  region where (see \fig{tsa}) $z >0$ and $z^+ > 0$, then
both amplitudes $N\left(r, R_0;Y'\right)$ and $N^+\left(R,r; Y - Y'\right)$ are deeply in the
saturation region. We can find the solution to \eq{EQ21} and \eq{EQ22} in this region,
using experience of solving the Balitsky-Kovchegov equation \cite{LTSOL}. In this approach $N$  is
replaced with $N = 1 +\Delta N$ (and similarly for $N^+$). This, together with the observation that constant $N^+$ in \eq{EQ21},
 and
constant $N$ in \eq{EQ22} don't contribute, we see that the asymptotic solution is
$N = 1$.   Replacing $N$ with $1+\De N$ in \eq{EQ21} and acting with $L_{xy}$ on both sides, and
using \eq{G01} leads to;

\beq \label{DEN}
 \frac{\partial \left(L_{xy}  \Delta N(x,y;Y') \right)}{\partial\,Y'}\,\,=
\eeq
$$
\,\,- \frac{\bas}{2
\pi}\,\int\,d^2\,z\,K\left(x,y|z\right)\, \left(L_{xy}  \Delta N(x,y;Y') \right)\,\,-\,\,
\frac{\bas}{
\pi}\,\int\,d^2\,z\,K\left(x,y|z\right)\left(L_{xz}  \Delta N^+(x,z;Y') \right)
$$
In \eq{DEN} we neglect all contributions of the order $(\Delta N)^2$, $ \Delta N\,\Delta N^+$ and
$(\Delta N^+)^2$.  A similar result is found by using the analogous above mentioned approach on \eq{EQ22}.

\FIGURE[h]{\centerline{\epsfig{file=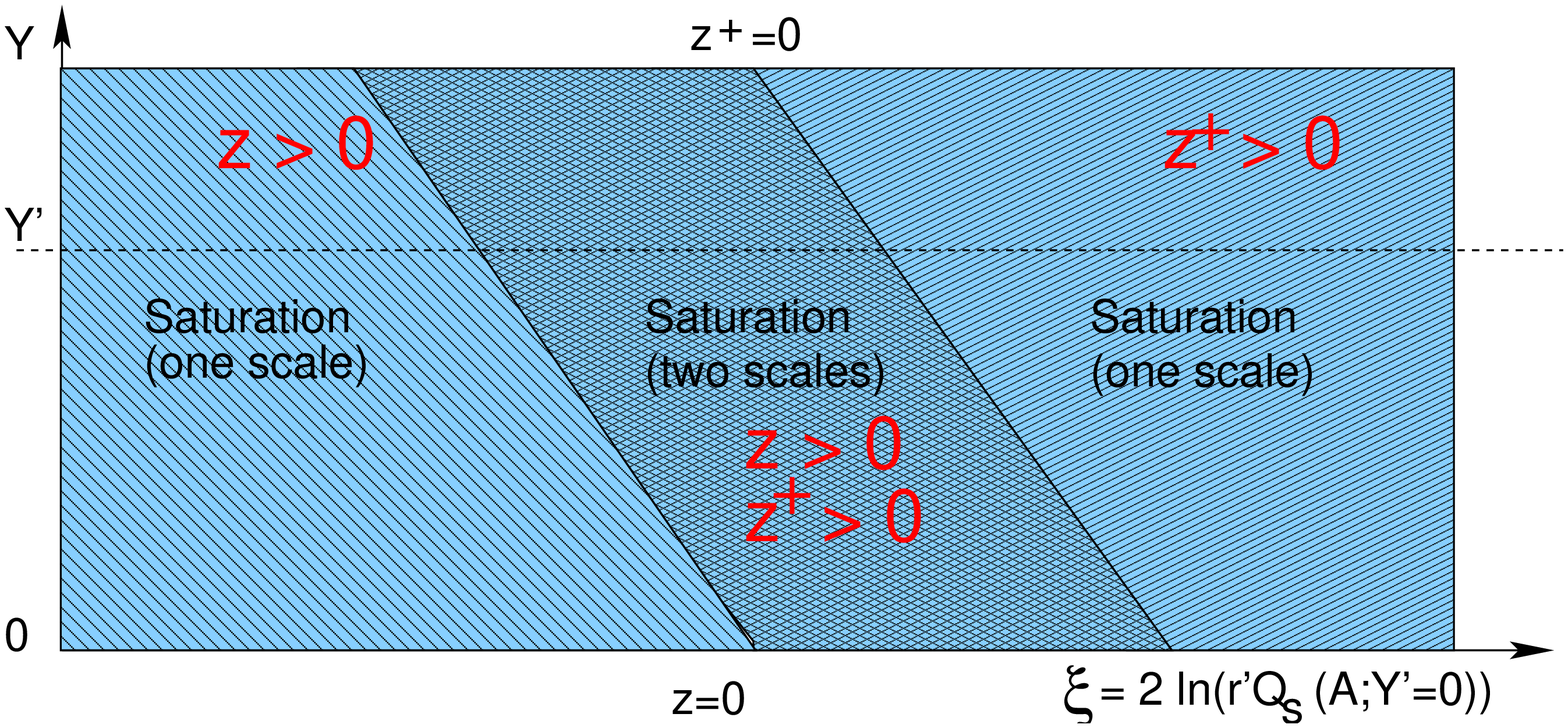,width=140mm}}
\caption{Two saturation scales for  nucleus-nucleus scattering.}\label{tsa}}


Introducing  the new functions $\tilde{n} = L_{x y} \Delta N$ and $\tilde{n}^+ \,=\,L_{x y} \Delta N^+$ we see that \eq{EQ21} and
 \eq{EQ22} reduce to two linear equations
\bea\frac{\partial  \tilde{n}\Lb x,y,Y'\Rb}{\partial\,Y'}\,\,&=&\,\,
\,\,- \frac{\bas}{2
\pi}\,\int\,d^2\,z\,K\left(x,y|z\right)\,\tilde{n}\Lb x,y,Y'\Rb \,\,-\,\,
\frac{\bas}{
\pi}\,\int\,d^2\,z\,K\left(x,y|z\right)\tilde{n}^+\Lb z,y,Y'\Rb\,\hspace{1cm}\label{SDS1010}\\
-\,\frac{\partial  \tilde{n}^+\Lb x,y,Y'\Rb}{\partial\,Y'}\,\,&=&\,\,
\,\,- \frac{\bas}{2
\pi}\,\int\,d^2\,z\,K\left(x,y|z\right)\,\tilde{n}^+\Lb x,y,Y'\Rb \,\,-\,\,
\frac{\bas}{
\pi}\,\int\,d^2\,z\,K\left(x,y|z\right)\tilde{n}\Lb z,y,Y'\Rb\,\hspace{1cm}\label{SDS1020}\eea

We first rewrite these equations in a more convenient form:

\bea
\frac{\partial  \tilde{n}\Lb x,y,Y'\Rb}{\partial\,Y'}=&& -\,\, \frac{\bas}{
2\pi}\,\int\,d^2\,z\,K\left(x,y|z\right)\,\tilde{n}\Lb x,y,Y'\Rb \,\,-\,\, \frac{\bas}{
2\pi}\,\int\,d^2\,z\,K\left(x,y|z\right)\,\tilde{n}^+\Lb x,y,Y'\Rb\hspace{1cm} \label{SDS201}\\
&&-
\frac{\bas}{2
\pi}\,\int\,d^2\,z\,K\left(x,y|z\right)\,\Big\{ 2 \tilde{n}^+\Lb x,z,Y'\Rb\,\,-\,\,\tilde{n}^+\Lb x,y,Y'\Rb \Big\} \nn\\
\nn\\
-\,\,\frac{\partial  \tilde{n}^+\Lb x,y,Y'\Rb}{\partial\,Y'}=&&
  -\,\, \frac{\bas}{
2\pi}\,\int\,d^2\,z\,K\left(x,y|z\right)\,\tilde{n}^+\Lb x,y,Y'\Rb \,\,-\,\, \frac{\bas}{
2\pi}\,\int\,d^2\,z\,K\left(x,y|z\right)\,\tilde{n}\Lb x,y,Y'\Rb\hspace{1cm}  \label{SDS202}\\
  &&-
\frac{\bas}{2
\pi}\,\int\,d^2\,z\,K\left(x,y|z\right)\,\Big\{ 2 \tilde{n}\Lb x,z,Y'\Rb\,\,-\,\,\tilde{n}\Lb x,y,Y'\Rb \Big\}\nn
\eea

We calculate $\int d^z K\Lb x, y|z\Rb$ taking into account that
the main contribution in the saturation  region stems from the decay of the large size dipole into one small size dipole  and one large size dipole. Indeed,
\bea
&& \int d^2 z \,K\Lb x_1, x_2| z\Rb\,\,\rightarrow\,\pi\,\int \frac{x^2_{12} d x_{13}^2}{x^2_{13}\,|x^2_{12} - x^2_{13}|}
\,\,=\,\,\pi\int^{x^2_{12}}_{\rho^2} \frac{d x^2_{13}}{x^2_{13}} \,\,+\,\,\pi\int^{x^2_{12}}_{\rho^2}\frac{d |x^2_{12} - x^2_{13}|}
{|x^2_{12} - x^2_{13}|}\label{LOGK}\\
&&=\,\,2 \pi \ln\Lb x^2_{12}/\rho^2\Rb \,\,=\,\,2\pi \,\xi\nn
\eea
It should be noted that the $\xi$ which is defined in \eq{LOGK}, is different from the $\xi$ that appears in \fig{ts} and \fig{tsa}.
At this stage we have introduced the artificial cutoff at small values of the dipole size ($\rho$),
but the physical cutoff is $\rho=1/Q_s$, as discussed in Ref.\cite{LTSOL}. It should be noticed that for the opposite case, when a dipole decays to two dipoles of larger sizes,
there is no logarithmic contribution, since

\beq \label{LOGK1}
\int d^2 z  \,K\Lb x_1, x_2|z \Rb\,\,\rightarrow\,\pi\, \int^{1/Q^2_s}_{x^2_{12}} \frac{ d x^2_{13}}{x'^4_{13}}\,\,
\eeq

In the two scale kinematic region, the size of the dipole is restricted as follows (see \fig{tsa} and \eq{TS41} and \eq{TS42});
\beq \label{SIZER}
       1/Q^2_{1,s} \,\,\,\geq\,\,\,r'^2 \,\,\,\geq\,\,\, 1/Q^2_{2,s}
\eeq
Therefore, the natural choice is $\rho^2 = 1/Q^2_{2,s}$. Using \eq{LOGK}, then
\eq{SDS201} and \eq{SDS202} simplify to the following;

\bea
\frac{\partial  \tilde{n}\Lb x,y,Y'\Rb}{\partial\,Y'}\,\,=&& \,\,
- \bas\,z^+\,\Big\{\tilde{n}\Lb x,y,Y'\Rb \,+ \,\tilde{n}^+\Lb x,y,Y'\Rb \Big\}\label{SDS301}\\
&&-\,\,
\frac{\bas}{2
\pi}\,\int\,d^2 z\,K\left(x,y|z\right)\,\Big\{ 2 \tilde{n}^+\Lb x,z,Y'\Rb\,\,-\,\,\tilde{n}^+\Lb x,y,Y'\Rb \Big\}\,\nn\\
\nn\\
-\,\frac{\partial  \tilde{n}^+\Lb x,y,Y'\Rb}{\partial\,Y'}\,\,=&&
 \,\,- \bas\,z^+\,\Big\{\tilde{n}\Lb x,y,Y'\Rb \,+ \,\tilde{n^+}\Lb x,y,Y'\Rb \Big\}\label{SDS302}\\
&&-\,\,
\frac{\bas}{2
\pi}\,\int\,d^2 z\,K\left(x,y|z\right)\,\Big\{ 2 \tilde{n}\Lb x,z,Y'\Rb\,\,-\,\,\tilde{n}\Lb x,y,Y'\Rb \Big\}\,\nn
\eea

We use the double Mellin transform to solve \eq{SDS301} and \eq{SDS302}, namely
\bea
\tilde{n}\Lb z^+,Y'\Rb\,\,\,&&=\,\,\int^{\epsilon + i \infty}_{\epsilon - i \infty}\,\frac{d \omega}{2 \pi i}\int^{\epsilon + i \infty}_{\epsilon - i \infty}\,\frac{d \gamma}{ 2 \pi i}\,\,\tilde{n}\Lb \omega, \gamma\Rb \,e^{ \omega Y' + (1 - \gamma)\,z^+}\label{SDS4}\\
\tilde{n}^+\Lb z,Y'\Rb\,\,\,&&=\,\,\int^{\epsilon + i \infty}_{\epsilon - i \infty}\,
\frac{d \omega}{2 \pi i}\int^{\epsilon + i \infty}_{\epsilon - i \infty}\,\frac{d \gamma}{ 2 \pi i}\,\,
\tilde{n}^+\Lb \omega, \gamma\Rb \,e^{ \omega Y' + (1 - \gamma)\,z^+}\label{SDS4a}
\eea

Substituting \eq{SDS4} and \eq{SDS4a} into \eq{SDS301} and \eq{SDS302}, adding and subtracting these two equations  we obtain for
$\Sigma\Big(\om,\ga\Big)\,=\,\tilde{n}\Lb \omega, \gamma\Rb\,+\,\tilde{n}^+\Lb \omega, \gamma\Rb$ and
$\Delta\Big(\om,\ga\Big)\,=\,\tilde{n}\Lb \omega, \gamma\Rb\,-\,\tilde{n}^+\Lb \omega, \gamma\Rb$ the following;

\bea
&&\Lb \omega + \lambda_s (1 - \gamma)\Rb \,\Delta\Lb \omega, \gamma\Rb\,\,=\,\,-\,2\,\bas\, \frac{\partial \Sigma\Lb \omega, \gamma\Rb}{\partial \gamma} \,\,-\,\,\omega\Lb \gamma\Rb \Sigma\Lb \omega, \gamma\Rb \label{SDS501}\\
&&\Lb \omega + \lambda_s (1 - \gamma)\Rb  \,\Sigma\Lb \omega, \gamma\Rb\,\,=\,\,\omega\Lb \gamma\Rb \Delta\Lb \omega, \gamma\Rb; \label{SDS502}
\eea

where $\omega\Lb \gamma\Rb $ is given by \eq{BFKLOM}. Solving \eq{SDS501} we find

\bea
\Sigma\Lb \omega, \gamma\Rb\,\,&&=\,\, \Sigma_0\Lb \gamma\Rb\,\exp\Big\{-\,\int^{\gamma}_0 \,d \gamma^{\,\prime}\,
\frac{\Big( \omega + \lambda_s (1 - \gamma^{\,\prime})\Big)^2 + \omega^2\Lb \gamma^{\,\prime}\Rb}{2 \bas\,\omega\Lb \gamma
^{\,\prime}\Rb}\Big\}\label{SDS6}\\
&&=\,\,\Sigma_0\Lb \gamma\Rb\,\exp\Big\{\,L\Lb\ga\Rb\om^2+F\Lb\ga\Rb\om+K\Lb\ga\Rb\Big\}\label{SDS6a}
\eea

where we define

\bea
L\Lb\ga\Rb&&=-\f{1}{2\bas}\int^\ga_0\f{d\ga^{\,\prime}}{\om\Lb\ga^{\,\prime}\Rb}\lab{L}\\
F\Lb\ga\Rb&&=-\f{1}{\bas}\int^\ga_0d\ga^{\,\prime}\f{\la_s\Lb 1-\ga^{\,\prime}\Rb}{\om\Lb\ga^{\,\prime}\Rb}\lab{Fg}\\
K\Lb\ga\Rb&&=-\f{1}{2\bas}\int^\ga_0d\ga^{\,\prime}\f{\la_s^2\Lb 1-\ga^{\,\prime}\Rb^2}{\om\Lb\ga^{\,\prime}\Rb}-\f{1}{2\bas}\int^\ga_0d\ga^{\,\prime}\om
\Lb\ga^{\,\prime}\Rb\lab{Kg}\eea

and where $\Sigma_0\Lb \gamma\Rb$ has to be found from the matching with the regions with one saturation scale (see \fig{tsa}).
Using \eq{SDS6} we obtain the solution in the form

\bea
\Sigma\Lb z^+,Y'\Rb\,\,\,&&=\,\,\int^{\epsilon + i \infty}_{\epsilon - i \infty}\,\frac{d \omega}{2 \pi i}\int^{\epsilon + i \infty}_{\epsilon - i \infty}\,\frac{d \gamma}{ 2 \pi i}\,\,\Sigma_0\Lb \ga\Rb \,e^{\Psi\Lb\omega,\gamma,Y',z^+\Rb}\label{SDS7}\\
\Psi\Lb\omega,\gamma,Y',z^+\Rb\, \,&&=\,\, \omega \Lb Y^{\,\prime}+F\Lb\ga\Rb\Rb \, + \,(1 - \gamma)\,z^+ +\om^2 L\Lb\ga\Rb \label{SDS8}
\eea

where the function $K\Lb\ga\Rb$ which appears in the exponential function in \eq{SDS6a} has been absorbed into the function $\Sigma_0\Lb\ga\Rb$ in \eq{SDS7}.
The integral over $\omega$ in \eq{SDS7} can be taken explicitly leading to the following expression
\bea
\Sigma\Lb z^+,Y'\Rb\,\,\,&&=\,-\,\int^{\epsilon + i \infty}_{\epsilon - i \infty}\,\frac{d \gamma}{ 2 \pi i}\,\,\Sigma_0\Lb \gamma\Rb \, \sqrt{\frac{\pi}{L\Lb \gamma\Rb}}\,\exp\Big( \,(1 - \gamma)\,z^+\,
 - \frac{\Lb Y'+F\Lb\ga\Rb\Rb^2}{4 L\Lb \gamma\Rb} \Big)\lab{SDS9a}\\
 &&=\,-\,\int^{\epsilon + i \infty}_{\epsilon - i \infty}\,\frac{d \gamma}{ 2 \pi i}\,\,\tilde{\Sigma}_0\Lb \gamma\Rb \, \,\exp\Big( \,(1 - \gamma)\,z^+\,
 - \f{Y^{\,\prime}}{4L\Lb\ga\Rb}\Lb Y^{\,\prime}+2F\Lb\ga\Rb\Rb\Big)\lab{SDS9}\eea

where passing from \eq{SDS9a} to \eq{SDS9}, a factor of $\Lb\pi/L\Lb\ga\Rb\Rb^{1/2}$ and $\exp\Lb-F^2\Lb\ga\Rb/4L\Lb\ga\Rb\Rb$ have been absorbed into the function
$\tilde{\Sigma}_0\Lb\ga\Rb$.
From \eq{SDS501} and \eq{SDS6a} we can deduce $\De\Lb\om,\ga\Rb$, viz;

\bea
\De\Lb\om,\ga\Rb&&=\f{\Lb \om+\la_s\Lb 1-\ga\Rb\Rb}{\om\Lb\ga\Rb} \Sigma_0\Lb \gamma\Rb\,\exp\Big\{\,L\Lb\ga\Rb\om^2+F\Lb\ga\Rb\om+K\Lb\ga\Rb\Big\}\lab{SDS10a}\eea

Using the double Mellin transform of \eq{SDS4} we can transform $\De\Lb\om,\ga\Rb$ to $\De\Lb z^+,Y^{\,\prime}\Rb$
which is a function of $z^+$ and $Y^{\,\prime}$ as

\bea
\De\Lb z^+,Y^{\,\prime}\Rb&&=\,\,\int^{\epsilon + i \infty}_{\epsilon - i \infty}\,\frac{d \omega}{2 \pi i}\int^{\epsilon + i \infty}_{\epsilon - i \infty}\,\frac{d \gamma}{ 2 \pi i}
\f{\Lb \om+\la_s\Lb 1-\ga\Rb\Rb}{\om\Lb\ga\Rb} \Sigma_0\Lb \gamma\Rb\,\lab{SDS101a}\\
&&\times \exp\Big\{\,\Lb 1-\ga\Rb z^++\Lb Y^{\,\prime}+F\Lb\ga\Rb\Rb\om+L\Lb\ga\Rb\om^2\Big\}\nn\eea

where a factor of $\it{\exp}\Lb K(\ga)\Rb$ has been absorbed into the function $\Sigma_0\Lb\ga\Rb$. The integration over $\om$ can be solved analytically, leading to

\bea
\De\Lb z^+,Y^{\,\prime}\Rb&&=\,-\,\int^{\epsilon + i \infty}_{\epsilon - i \infty}\,\frac{d \gamma}{ 2 \pi i} \Sigma_0\Lb \gamma\Rb\sqrt{\f{\pi}{L\Lb\ga\Rb}}
\Lb-\f{Y^{\,\prime}+F\Lb\ga\Rb}{2\om\Lb\ga\Rb L\Lb\ga\Rb}+\f{\la_s\Lb 1-\ga\Rb}{\om\Lb\ga\Rb}\Rb \, \lab{SDS10b}\\&&\exp\Big\{\,\Lb 1-\ga\Rb z^+-\f{\Lb Y^{\,\prime}+F\Lb\ga\Rb\Rb^2}{4L\Lb\ga\Rb}\Big\}\nn\\
\nn\\
&&=\,-\,\int^{\epsilon + i \infty}_{\epsilon - i \infty}\,\frac{d \gamma}{ 2 \pi i} \tilde{\Sigma}_0\Lb \gamma\Rb
\Lb-\f{Y^{\,\prime}+F\Lb\ga\Rb}{2\om\Lb\ga\Rb L\Lb\ga\Rb}+\f{\la_s\Lb 1-\ga\Rb}{\om\Lb\ga\Rb}\Rb\,\lab{SDS10c}\\&&\exp\Big\{\,\Lb 1-\ga\Rb z^+-\f{\,Y^{\,\prime}\,\Lb Y^{\,\prime}+2F\Lb\ga\Rb\Rb}{4L\Lb\ga\Rb}\Big\}\nn
\eea

where in passing from \eq{SDS10b} to \eq{SDS10c}, a factor of $\Lb\pi/L\Lb\ga\Rb\Rb^{1/2}$ and $\exp\Lb -F^2\Lb\ga\Rb/4L\Lb\ga\Rb\Rb$ has been absorbed into the function $\tilde{\Sigma}_0\Lb\ga\Rb$. From the definition
$\Sigma\Lb z^+, Y^{\,\prime}\Rb\,=\,\tilde{n}\Lb z^+, Y^{\,\prime}\Rb\,+\,\tilde{n}^+\Lb z^+, Y^{\,\prime}\Rb$ and
$\Delta\Lb z^+, Y^{\,\prime}\Rb\,=\,\tilde{n}\Lb z^+, Y^{\,\prime}\Rb\,-\,\tilde{n}^+\Lb z^+, Y^{\,\prime}\Rb$, then using \eq{SDS9} and \eq{SDS10c}
 we can obtain $\tilde{n}\Lb z^+,Y^{\,\prime}\Rb$ and $\tilde{n}^+\Lb z^+,Y^{\,\prime}\Rb$ as

\bea
\tilde{n}\Lb z^+, Y'\Rb\,\,&&=\,\,\int^{\epsilon + i \infty}_{\epsilon - i \infty}\,\frac{d \gamma}{ 2 \pi i}\,\,\tilde{\Sigma}_0\Lb \gamma\Rb
\h\,\Big( 1 +\theta\Lb \ga, Y^{\,\prime}\Rb\Big)\,
 \, \,\exp\Big( \,\Lb 1-\ga\Rb z^+
+Y^{\,\prime}\tau\Lb Y^{\,\prime},\ga\Rb\ \,\Big)\hspace{0.7cm}\label{SDS101}\\
\tilde{n}^+\Lb  z^+, Y'\Rb\,\,&&=\,\,\int^{\epsilon + i \infty}_{\epsilon - i \infty}\,\frac{d \gamma}{ 2 \pi i}\,\,\tilde{\Sigma}_0\Lb \gamma\Rb
\h\,\Big( 1 -\theta\Lb \ga, Y^{\,\prime}\Rb\Big)\,
\, \,\exp\Big( \,\Lb 1-\ga\Rb z^+
+Y^{\,\prime}\tau\Lb Y^{\,\prime},\ga\Rb
\ \,\Big) \hspace{0.7cm}
\label{SDS102}\eea

where we define

\bea
\theta\Lb\ga,Y^{\,\prime}\Rb&&=-\f{Y^{\,\prime}+F\Lb\ga\Rb}{2\om\Lb\ga\Rb L\Lb\ga\Rb}+\f{\la_s\Lb 1-\ga\Rb}{\om\Lb\ga\Rb}
=-\bas L\Lb\ga\Rb\f{d}{d\ga}\Big(\f{Y^{\,\prime}+F\Lb\ga\Rb}{L\Lb\ga\Rb}\Big)\lab{theta}\\
\tau\Lb\ga,Y^{\,\prime}\Rb&&=-\,\f{\Lb Y^{\,\prime}+2F\Lb\ga\Rb\,\Rb}{4L\Lb\ga\Rb}\lab{tau}\eea

where we have absorbed all factors that depend on $\gamma$ in the function $\tilde{\Sigma}_0(\gamma)$, which should be found from the boundary conditions.
For  $z^+=0$ the solution reduces to

\beq\lab{SDS11}
\tilde{n}\Lb z^+=0, Y' - Y_0\Rb=\int^{\epsilon + i \infty}_{\epsilon - i \infty}\,\frac{d \gamma}{ 2 \pi i}\tilde{\Sigma}_0\Lb \gamma\Rb
\h\Big( 1 +\theta\Lb \ga, Y^{\,\prime}-Y_0\Rb\Big)\exp\Big( \Lb Y^{\prime}-Y_0\Rb\tau\Lb\ga,Y^{\prime}-Y_0\Rb\Big)\eeq

In \eq{SDS11} we introduced the initial value of rapidity $Y=Y_0$ which was assumed to be equal  to zero above. Let us find the Green function demanding that $\tilde{n}\Lb z^+=0, Y' - Y_0\Rb\,\,=\,\,\delta \Lb Y' - Y_0\Rb$.
If we find such a solution, then integrating it over $Y_0$ with an arbitrary function will lead to any boundary
 condition. Actually, as we know at $z^+=0$, then $ \tilde{n} = Const$ (see Ref.\cite{LTSOL}) but we will discuss this condition below in more detail.
Choosing

\bea \tilde{\Sigma}_0\Lb \ga\Rb=\f{\Sigma_0}{\bas L\Lb\ga\Rb}\lab{Sigma0}\eea

we can rewrite the pre-exponential factor in \eq{SDS11} in the form

\bea
&&\tilde{\Sigma}_0\Lb \gamma\Rb
\h\,\Big( 1 + \theta\Lb\ga,Y^{\,\prime}\Rb\Big)\,\,\to\,\,\Sigma_0\Big(\f{1}{2\bas L\Lb\ga\Rb}+\h
  \frac{d \tau\Lb\ga,Y^{\,\prime}\Rb}{d \gamma}\,\,-\,\,\f{1}{4}\f{d}{d\ga}\Lb\f{F\Lb\ga\Rb}{L\Lb\ga\Rb}\Rb
+h\Lb\ga\Rb\Big) \label{SDS12}
\eea

where $h\Lb\ga\Rb$ is another arbitrary function of $\ga$ to be determined by boundary conditions.
It is clear that the $d \tau\Lb\ga,Y^{\,\prime}\Rb/d \ga$ term in \eq{SDS12} leads to $\delta\Lb Y' - Y_0\Rb$
after integration over $\ga$, since we have the freedom to change the integration variable in \eq{SDS11}
 to $\tau= -(Y' - Y_0+2F\Lb\ga\Rb)/(4\,L\Lb \gamma\Rb)$.
The second term as well as $h\Lb \ga\Rb$ gives the function that falls down as $e^{ - Y'^{3/2}}$ at large $Y'$.
Finally, the solution to \eq{SDS101} with the boundary condition $ \tilde{n}\Lb z^+=0, Y' - Y_0\Rb = \delta\Lb Y' - Y_0\Rb$ has the form

\beq \label{SDS13}
\tilde{n}\Lb z^+, Y'-Y_0\Rb\,\,=\,\,\Sigma_0\,\int^{\epsilon + i \infty}_{\epsilon - i \infty}\,\frac{d \gamma}{ 2 \pi i}\,\,
\Big\{\f{d\tau\Lb\ga\Rb}{d\ga}+\tilde{h}\Lb\ga\Rb\Big\}\,
 \, \,\exp\Big( \,(1 - \gamma)\,z^+\, +\Lb Y^{\,\prime}-Y_0\Rb\tau\Lb\ga,Y^{\,\prime}-Y_0\Rb \,\Big)
\eeq
where $\Sigma_0$ is a constant with respect to $Y'$ and $\xi$ and
\beq \label{TILDEH}
 \tilde{h}\Lb\ga\Rb\,\,=\,\,\f{1}{2\bas L\Lb\ga\Rb}
  \,\,-\,\,\f{1}{4}\f{d}{d\ga}\Lb\f{F\Lb\ga\Rb}{L\Lb\ga\Rb}\Rb
+h\Lb\ga\Rb
\eeq
The next step is to satisfy to the initial condition for $Y'=Y_0$.  Along this line the initial condition is given by \eq{TAU}, namely,\footnote{We hope that the notation $\zeta_0\Lb b \Rb$ that we use below, will not be confused with $\zeta$ in \eq{TS6}.}
\beq \label{SDS14}
\tilde{n}\Lb \xi, Y'=Y_0\Rb \,\,\,=\,\,\frac{2 \as^2 C_F}{N_c}\,x^2_{12}\,\ln\Lb x^2_{12}/r^2\Rb\,S_A\Lb b\Rb\,\,=\,\,
\frac{2 \as^2 C_F}{N_c}\,\frac{S_A\Lb b \Rb}{Q^2_s\Lb A,Y'=Y_0\Rb}\,\xi \,e^\xi\,\,=\,\,\zeta_0(b)\,\xi e^\xi
\eeq

Note that when $Y^{\,\prime}=Y_0$, then \eq{tau} implies that
  $d\tau\Lb \ga,Y^{\,\prime}=Y_0\Rb/d\ga=-d/d\ga\left\{ F\Lb\ga\Rb/2L\Lb\ga\Rb\right\}$.
Therefore, we need to find $\Sigma_0$ from the condition
\bea
\tilde{n}\Lb \xi, Y'=Y_0\Rb&=&\Sigma_0\int^{\epsilon + i \infty}_{\epsilon - i \infty}\frac{d \gamma}{ 2 \pi i}
 \Lb \f{1}{2\bas L\Lb\ga\Rb} - \f{1}{2}\f{d}{d\ga}\Lb\f{F\Lb\ga\Rb}{L\Lb\ga\Rb}\Rb
 + h\Lb\ga\Rb\Rb e^{(1 - \gamma)\xi} = \zeta_0(b)\xi e^\xi\hspace{3mm}\Rightarrow\nn\\
\nn\\
\tilde{n}\Lb \xi, Y'=Y_0\Rb&=&\Sigma_0\int^{\epsilon + i \infty}_{\epsilon - i \infty}\frac{d \gamma}{ 2 \pi i}\Lb \f{1}{2\bas L\Lb\ga\Rb}-\f{F\Lb\ga\Rb}{4\bas L^2\Lb\ga\Rb\om\Lb\ga\Rb} + \f{\la_s\Lb 1-\ga\Rb}{2\bas\om\Lb\ga\Rb L\Lb\ga\Rb} + h\Lb\ga\Rb
\Rb e^{(1 - \gamma)\xi}\hspace{1cm}\label{SDS15}\\
&=&\zeta_0(b)\xi e^\xi \nn
\eea

At $\ga \,\to\,0$, $F\Lb \ga \Rb \,\to\,2\,\lambda_s\,L\Lb \ga\Rb \,+\,{\cal O}\Lb \gamma^3\Rb$ and therefore,
the only singularity $1/\ga^2$ stems from the  $1/L\Lb\ga\Rb$ term in \eq{SDS15}, (assuming that $h(\ga)$ has no such singularity).
Therefore, we need to find $\Sigma_0$ from the condition
\beq \label{SDS151}
\tilde{n}\Lb \xi, Y'=Y_0\Rb\,\,=\,\,\Sigma_0\,\int^{\epsilon + i \infty}_{\epsilon - i \infty}\,\frac{d \gamma}{ 2 \pi i}\,\,
\frac{1}{2\bas L\Lb \gamma\Rb}\,\,e^{(1 - \gamma)\xi}\,\,\,=\,\,\,\zeta_0(b)\,\xi e^\xi
\eeq
Assuming that $\gamma$ in \eq{SDS151} will be small we see that (see \eq{L})
\beq \label{LS}
L\Lb \gamma\Rb \,\,=
\,\,\int^\gamma_0 d \gamma' \frac{1}{2\,\bas \omega\Lb \gamma'\Rb}\,\,\xrightarrow{\gamma \ll 1}\,\,\frac{1}{4 \bas^2}\, \gamma^2
\eeq
Using \eq{LS}, one can see that \eq{SDS15} is satisfied if;
\beq \label{SIG0}
\Sigma_0\,\,=\,\,\zeta_0(b)/2 \,\bas\,
\eeq

Therefore, we can consider $h\Lb \ga\Rb = 0$ if we  are not interested in the correction of the order of $1/\xi$ to the
initial condition of
\eq{SDS14}.
Finally, the solution which takes into account both initial and boundary conditions takes the form

\bea
\tilde{n}\Lb z^+, Y'-Y_0\Rb\,\,&&=\,\,\f{\zeta_0\Lb b\Rb}{4\bas^2}
\,\int^{\epsilon + i \infty}_{\epsilon - i \infty}\,\frac{d \gamma}{ 2 \pi i}
\,\, \frac{1}{L\Lb \ga \Rb}\,
\Big\{ 1\,\,+\,\,\theta\Lb \ga, Y'\Rb\,\Big\}\label{SDS16}\\
&&\times
 \, \,\exp\Big( \,(1 - \gamma)\,z^+\, - \Lb Y'  - Y_0\Rb\f{\Lb Y^{\,\prime}-Y_0+2F\Lb\ga\Rb\Rb}{4L\Lb\ga\Rb} \,\Big)
\nn\\
\nn\\
&& = \,\,\f{\zeta_0\Lb b\Rb}{4\bas^2}\,\int^{\epsilon + i \infty}_{\epsilon - i \infty}\,\frac{d \gamma}{ 2 \pi i}\,\, \frac{1}{L\Lb \ga \Rb}\,
\Big\{ 1\,\,+\,\,\theta\Lb \ga, Y'\Rb\,\Big\}\,e^{\Psi\Lb Y',z^+,\ga \Rb}\nn
\eea

In order to find $\tilde{n}\Lb z, Y-Y^{\,\prime}\Rb$, we notice that \eq{SDS101}  at  the rapidity $Y - Y'$ is also a solution to the equation. Using the fact that from \eq{TS5} $z^+=\la_sY -z$, then  \eq{SDS16}  can be recast  as

\bea
\tilde{n}\Lb z^+, Y-Y'\Rb\,\,&&=\,\,\Sigma_0\Lb Y \Rb\,\int^{\epsilon + i \infty}_{\epsilon - i \infty}\,\frac{d \gamma}{ 2 \pi i}\,\,
\frac{1}{2\bas L\Lb \ga \Rb}\,
\Big\{ 1\,\,+\,\,\theta\Lb \ga, Y -  Y'\Rb\,\Big\}\label{SDS161}\\
&&\times
 \, \,\exp\Big( \,(1 - \gamma)\,\Lb \la_s Y \,-\,z\Rb \, - \Lb Y - Y'\Rb\f{\Lb Y-Y'+2F\Lb\ga\Rb\Rb}{4L\Lb\ga\Rb} \,\Big)
\nn\eea
However generally speaking,
$\Sigma_0$ in \eq{SDS161}  is not a constant, but rather a function of $Y$. Using \eq{SIG0} and by choosing this function  to take the form
\beq \label{SIG01}
\Sigma_0\Lb Y \Rb\,\,=\,\,\Sigma_0\Lb; \vec{B} - \vec{b} \Rb \,\exp\Lb- \la_s \,Y\Rb\,\,=\,\,
\f{\zeta_0\Lb \vec{B} - \vec{b} \Rb}{2\bas} \,\exp\Lb- \la_s \,Y\Rb\eeq
we obtain the required solution for $\tilde{n}^+\Lb z, Y - Y'\Rb$. Finally

\bea
\tilde{n}^+\Lb z, Y - Y'\Rb\,\,&&=\,\,\f{\zeta_0\Lb \vec{B}-\vec{b}\Rb}{4\bas^2}
\,\int^{\epsilon + i \infty}_{\epsilon - i \infty}\,\frac{d \gamma}{ 2 \pi i}\,\, \frac{1}{L\Lb \ga \Rb}\,
\Big\{1\,\,+\,\,\theta\Lb \ga, Y -  Y'\Rb\Big\}\label{SDS162}\\
&&\times
 \, \,\exp\Big( \,-\,(1 - \gamma)\,z \, - \Lb Y - Y'\Rb\f{\Lb Y-Y'+2F\Lb\ga\Rb\Rb}{4L\Lb\ga\Rb} \,\Big)
\nn\eea

Using the steepest decent method we can estimate the typical value of $\ga$ in the integral of \eq{SDS16}.
The saddle point equation looks as follows
\beq \label{SDLPO}
\frac{\partial\,\Psi\Lb Y',z^+,\ga \Rb}{\partial \ga}\,=\,- z^+ \,-\,\frac{\Lb Y' - Y_0\Rb^2}{4\,L^2\Lb \ga_{SP}\Rb}\,L'_\ga\Lb \ga_{SP}\Rb
\,-\,\Lb Y' - Y_0\Rb\,\Lb \frac{2 F\Lb \ga_{SP}\Rb}{4 \,L\Lb \ga_{SP}\Rb}\Rb'_\ga\,\,=\,\,0
\eeq
The solution to \eq{SDLPO} leads to larger values of $\ga_{SP}$. Indeed, at large $\ga$ we have
\beq \label{LAGA}
\omega\Lb \gamma\Rb \,\,\xrightarrow{ \gamma = i \kappa; \,\,\kappa \gg 1}\,\,-\,2\,\bas\,\ln ( \kappa);\,\,
L\Lb  \gamma\Rb\,\,\xrightarrow{ \gamma = i \kappa; \,\,\kappa \gg 1}\,\,\frac{1}{4 \bas^2}\,\f{\gamma}{\ln\Lb \kappa\Rb}\,;
\,\,F\Lb  \gamma\Rb\,\,\xrightarrow{ \gamma = i \kappa; \,\,\kappa \gg 1}\,\,-\,\frac{\la_s}{4\, \bas^2}\,\f{\ga^2}{\ln\Lb \kappa\Rb}
\eeq
After substituting these expressions into \eq{SDLPO} we obtain that
\beq \label{SDLPO1}
\ga_{SP}\,\,\,=\,\,\,\sqrt{\frac{\Lb Y' - Y_0\Rb^2 \bas^2\,\ln(\kappa)}{z^+\,-\,\la_s\,\Lb Y' - Y_0\Rb/2}}
\eeq
and the solution behaves as
\bea \label{SDS17}
\tilde{n}\Lb z^+, Y' - Y_0\Rb  &\,\,\,\propto\,\,\,& \,H\Lb Y' - Y_0, z^+\Rb\,\exp\Lb\,\,z^+\,\, -\,\,{\cal Z}\Rb \\
&\mbox{where}&\,\,\,\,\,\,\,\,\,\,
 {\cal Z}\,=\,\sqrt{4\,\Lb Y' - Y_0\Rb^2 \bas^2\,\ln(\tilde{\kappa})\,\Lb z^+\,-\,\la_s\,\Lb Y' - Y_0\Rb/2\Rb} \nn\\
 &\mbox{and}& \,\,\,\,\,\,\,\,\,\,\tilde{\kappa}\,=\,\sqrt{\frac{\Lb Y' - Y_0\Rb^2 \bas^2}{z^+\,-\,\la_s\,\Lb Y' - Y_0\Rb/2}}\nn
\eea
where $H$ is a function which varies slowly with $Y' - Y_0$ and $z^+$. Using the same approach we obtain for $ \tilde{n}^+\Lb z, Y - Y'\Rb$ the following expression
\bea \label{SDS171}
\tilde{n}^+\Lb z, Y - Y'\Rb  &\,\,\,\propto\,\,\,& \,H^+\Lb Y - Y', z\Rb\,\exp\Lb\,\,-\,z\,\, -\,\,{\cal Z}^+\Rb \\
 &\mbox{where}&\,\,\,\,\,\,\,\,\,\,
 {\cal Z}^+\,=\,\sqrt{4\,\Lb Y - Y'\Rb^2 \bas^2\,\ln(\tilde{\kappa}^+)\,\Lb - z\,+\,\la_s\,\Lb Y - Y'\Rb/2\Rb} \nn\\
 &\mbox{and}& \,\,\,\,\,\,\,\,\,\,\tilde{\kappa}^+\,=\,\sqrt{\frac{\Lb Y - Y'\Rb^2 \bas^2}{\,z\,-\,\la_s\,\Lb Y - Y'\Rb/2}}\nn
\eea

Replacing $\ln\Lb \kappa\Rb$ by $ \ln\Lb \tilde{\kappa}\Rb$ one can calculate the pre-exponential
factors in \eq{SDS17} and \eq{SDS171} using \eq{LAGA} and the formula {\bf 8.432(6)} of Ref. \cite{RY}, namely
\bea\label{SDS181}
\tilde{n}\Lb \cal Z\Rb\,&=&\,2\zeta(b)\,\ln\Lb \tilde{\kappa}\Rb\,e^{z^+}
\,\Big\{\, \bas\Lb Y^{\,\prime}-Y_0\Rb\,\f{\Lb z^+-\la_s\Lb Y^{\,\prime}-Y_0\Rb/2\Rb}{\cal{Z}}
K_1\Lb \cal Z\Rb\,+\,\,K_0\Lb \cal Z\Rb\Big\}\\
\label{SDS182}\tilde{n}^+\Lb \cal Z^+\Rb\,&=&\,2\zeta(\vec{B} - \vec{b})
\ln\Lb \tilde{\kappa}^+\Rb\,e^{-z}
\,\Big\{ \bas \Lb Y-Y^{\,\prime}\Rb\,\f{\Lb- z+\la_s\Lb Y- Y^{\prime}\Rb/2\Rb}{\cal{Z^+}}
K_1\Lb \cal Z^+\Rb\,+\,K_0\Lb \cal Z^+\Rb\Big\}
\hspace{1cm}\eea

\subsection{Solution in the region with one saturation scale}
The solution given by \eq{SDS17}, is correct in the region with two saturation scales, or in other words,
the region between the two lines $z^+=0$ and $z=0$ in \fig{tsa}.  In the region to the right of the line $z=0$, then $z^+$ becomes negative and $\tilde{n}^+$ is small here.
Therefore, the equation for $\tilde{n}$ takes the form
\beq\label{SOS1}
\omega\tilde{n}\Lb\omega,\gamma\Rb
\,\,\,=\,\,\,- \bas\,\frac{\partial\tilde{n}\Lb \omega,\gamma\Rb}{\partial \,\gamma}
\eeq

instead of \eq{SDS501} and \eq{SDS502}. The solution to this equation is very simple
\beq \label{SOS2}
\tilde{n}\Lb \omega,\gamma\Rb\,\,\,=\,\,\, \tilde{n}_0\Lb \gamma\Rb\,\exp \Big\{ - \bas\,\omega\,\gamma\Big\}
\eeq
where we need to find $\tilde{n}\Lb \omega,\gamma\Rb$ from the matching of this solution with the solution in the kinematic region
 with two saturation scales at $z=0$ (see \eq{SDS17}). Substituting \eq{SOS2} into \eq{SDS4} we obtain
\beq \label{SOS3}
\tilde{n}\Lb z^+,Y'\Rb\,\,\,=\,\,\int^{\epsilon + i \infty}_{\epsilon - i \infty}\,\frac{d \omega}{2 \pi i}\int^{\epsilon + i \infty}_{\epsilon - i \infty}\,\frac{d \gamma}{ 2 \pi i}\,\,\tilde{n}_0\Lb \gamma; Y\Rb \,e^{ \omega Y ' + (1 - \gamma)\,z^+ -\,\,\bas \,\omega\,\gamma}
\eeq
Integrating over $\omega$ we see that $\bas \gamma = Y'$ and finally we have
\beq \label{SOS4}
\tilde{n}\Lb z^+,Y' \Rb\,\,\,=\,\,\tilde{n}_0\Lb Y'; Y\Rb e^{  z^+ \,\,-\,\, \Lb Y' - Y_0\Rb \,z^+/\bas}
\eeq
which coincides with the solution given in
Ref.\cite{LTA}. At $z=0$, then $z^+= \zeta$, \vspace{1mm}\\ ${\cal Z} = \sqrt{\frac{\bas^2\ln\Lb (Y' - Y_0)/\sqrt{\zeta - \la_sY'/2}\Rb}{\zeta  - \la _sY'/2}\,\Lb Y' - Y_0\Rb^2}$,  and keeping only the factors in the exponent, we obtain
\beq \label{SOS5}
 \tilde{n}_0\Lb Y'; Y\Rb\,\,=\,\,e^{\eta\Lb Y',Y\Rb}\,\,\,
\eeq
with
\beq \label{SOS6}
\eta\Lb Y',Y\Rb\,\,=\,\,-\,\zeta\,\,+
 \,\,\,(Y'\,-\,Y_0)\,\zeta/\bas\,+\,\sqrt{\frac{\bas^2\ln\Lb (Y'\, -\, Y_0)/\sqrt{\zeta - \la_s (Y' - Y_0)}\Rb}{\zeta - \la_s (Y' - Y_0)}\,\Lb Y' - Y_0\Rb^2}
\eeq

Therefore, the answer in the kinematic region with $z < 0$, is
\beq \label{SOS7}
\tilde{n}\Lb z^+ ,Y',Y| z^+\,<\,0\Rb\,
\,\,=\, \,H\Lb \mbox{smooth function of $z^+$ and $ Y'$}\Rb\,\,e^{  \eta\Lb Y',Y\Rb + z^+ - z^+
\Lb Y' - Y_0\Rb/\bas}
\eeq

It is easy to see that the solution for $ \tilde{n}^+\Lb z ,Y',Y| z<0\Rb$ can be obtained from \eq{SOS7}  with the replacement:
 $Y'\, \to\, Y - Y'$ and $z^+\,\to\,z$ which yields;

\beq \label{SOS8}
\tilde{n}\Lb z ,Y',Y| z\,<\,0\Rb\,
\,\,=\, \,H\Lb \mbox{smooth function of $z$ and $Y\,-\, Y'$}\Rb\,\,e^{  \eta^+\Lb Y',Y| \eq{SOS9}\Rb\,\,+\,\,z\,\,-\,\,z\,\Lb Y\,- \,Y'\Rb}/\bas
\eeq
with
\beq \label{SOS9}
\eta^+\Lb Y',Y\Rb\,\,=\,\,-\,\zeta\,\,+
 \,\,\,\Lb Y\,-\,Y'\Rb \,\zeta/\bas\,+\,\sqrt{\frac{\bas^2\ln\Lb(Y\, -\,Y')/\sqrt{\la_s(Y - Y') - \zeta}\Rb}{\zeta \,-\,\la_s ( Y - Y')}\,\Lb Y\, - \,Y'\Rb^2}
\eeq

It should be stressed that the variable $\xi(b)$ that we use in the definition  of $z^+$ and $z$  is equal to
\bea \label{XIPL}
\xi\Lb\vec{b} \Rb \,\,=\,\,\ln\Big( r'^2\, \,S_{A_1} \Lb  \vec{b} \Rb\int d^2 b' \,Q^2_{1,s}\Lb \mbox{proton},\,Y'\,= \,0;  b'\Rb\Big) & & \mbox{for}\,\,\,z^+\\
\xi\Lb\vec{B}\, - \, \vec{b} \Rb \,\,=\,\,\ln\Big( r'^2\, \,S_{A_2} \Lb \vec{B}\, - \, \vec{b} \Rb\int d^2 b' \,Q^2_{2,s}\Lb \mbox{proton},  Y\,-\,Y'\,= \,0;  b'\Rb\Big) & & \mbox{for}\,\,\,z\nn
\eea

\section{The nucleus-nucleus scattering amplitude  deeply in the saturation region}

Recalling  that $\tilde{n}\,=\,L_{xy}\Delta N$ and  $\tilde{n}^+\,=\,L_{xy}\Delta N^+$ we first need to find  $\Delta N$ and $\Delta N^+$. Using \eq{LG} and \eq{SDS16}  we can  see that
\bea
\Delta N \Lb z^+, Y'\Rb\,\,\,\,\,\,\,\,\,\,&=& \,\,\f{\zeta\Lb b\Rb}{4\bas^2}\,\int^{\epsilon + i \infty}_{\epsilon - i \infty}\,
\frac{d \gamma}{ 2 \pi i}\,\, \lambda\Lb 0,\nu\Rb\,\frac{1}{L\Lb \ga \Rb}\,
\Big\{ 1\,\,+\,\,\theta\Lb \ga, Y'\Rb\,\Big\}\label{AM101}\\
&\times &
 \, \,\exp\Big( \,(1 - \gamma)\,z^+\, - \Lb Y'  - Y_0\Rb\f{\Lb Y^{\,\prime}-Y_0+2F\Lb\ga\Rb\Rb}{4L\Lb\ga\Rb} \,\Big)
\nn\\
\nn\\
\nn\\
 &=&\,\,\f{\zeta\Lb b\Rb}{4\bas^2}\,\int^{\epsilon + i \infty}_{\epsilon - i \infty}\,
\frac{d \gamma}{ 2 \pi i}\,\, \frac{1}{16 \gamma^2\,(1 - \gamma)^2}\,\frac{1}{L\Lb \ga \Rb}\,
\Big\{ 1\,\,+\,\,\theta\Lb \ga, Y'\Rb\,\Big\}\nn\\
&\times&
 \, \,\exp\Big( \,(1 - \gamma)\,z^+\, - \Lb Y'  - Y_0\Rb\f{\Lb Y^{\,\prime}-Y_0+2F\Lb\ga\Rb\Rb}{4L\Lb\ga\Rb} \,\Big)
\nn \\
\nn\\
\nn\\
\Delta N^+ \Lb z, Y - Y'\Rb &=& \,\,\f{\zeta\Lb  \vec{B} - \vec{b} \Rb}{4\bas^2}
\,\int^{\epsilon + i \infty}_{\epsilon - i \infty}\,\frac{d \gamma}{ 2 \pi i}\,\,\frac{1}{16 \gamma^2\,(1 - \gamma)^2} \frac{1}{L\Lb \ga \Rb}\,
\Big\{ 1\,\,+\,\,\theta\Lb \ga, Y -  Y'\Rb\,\Big\}\label{AM102}\\
&\times &
 \, \,\exp\Big( \,(1 - \gamma)\,\,z\, - \Lb Y - Y'\Rb\f{\Lb Y-Y'+2F\Lb\ga\Rb\Rb}{4L\Lb\ga\Rb} \,\Big)
\nn
\eea
Assuming that large values of $\gamma$ contribute to the integrals in \eq{AM101} and \eq{AM102},
we can rewrite these equations in a more economical format, after using the formulae {\bf 3.471(9)} and {\bf 8.432(6)} in Ref.\cite{RY}.
In this approach, after integrating over $\ga$, then \eq{AM101} and \eq{AM102} lead to the following expressions for
 $\De N$ and $\De N^+$
\bea
\Delta N \Lb z^+, Y'\Rb\,&&=\,\,2\zeta(b)\,\ln\Lb \tilde{\kappa}\Rb\,\,e^{z^+}\,\Big\{
\,\Lb\f{z^+-\la_s\Lb Y^{\,\prime}-Y_0\Rb/2}{{\cal Z}}\Rb^4\,K_4\Lb \cal Z\Rb \label{AM201}\\
&& +\,\,
2\bas\Lb Y^{\,\prime}-Y_0\Rb\Lb\f{z^+-\la_s\Lb Y^{\,\prime}-Y_0\Rb/2}{{\cal Z}}\Rb^5 K_5\Lb \cal Z\Rb
\Big\}
  \nn\\
\nn\\
\nn\\
\!\!\Delta N ^+\Lb z,Y\,-\, Y'\Rb\,&&=
\,\,2\zeta(\vec{B}-\vec{b})\,\ln\Lb \tilde{\kappa}^+\Rb\,\,e^{-z}\,\Big\{
\,\Lb\f{-z+\la_s\Lb Y-Y^{\,\prime}\Rb/2}{{\cal Z}^+}\Rb^4\,K_4\Lb {\cal Z}^+\Rb \label{AM202}\\\
&& +\,\,
2\,
\bas\Lb Y- Y^{\,\prime}\Rb\Lb\f{-z+\la_s\Lb Y- Y^{\,\prime}\Rb/2}{{\cal Z}^+}\Rb^5 K_5\Lb {\cal Z}^+\Rb\Big\}
  \nn\eea


\FIGURE[ht]{
\centerline{\epsfig{file=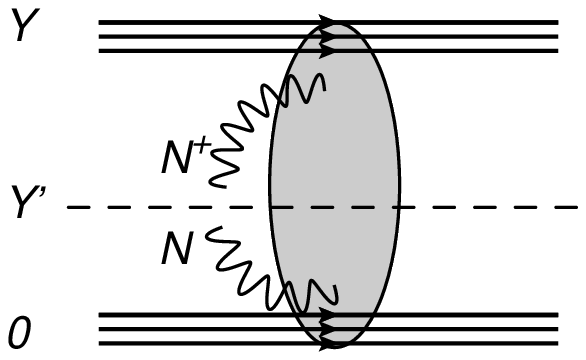,width=70mm}}
\caption{The graphical representation of the equation used for calculating the scattering amplitude using the $t$-channel unitarity.}
\label{calam}
}

\FIGURE[ht]{
\centerline{\epsfig{file=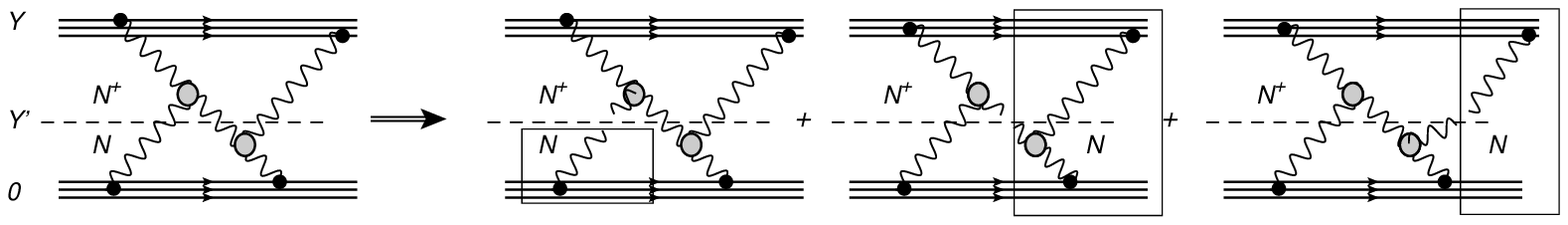,width=180mm}}
\caption{The example of the diagram which is described in terms of \protect\eq{AM4}.}
\label{calam1}
}



In \eq{AM201} and \eq{AM202} we use the same notation convention that was used in \eq{SDS181} and \eq{SDS182}. The obvious way to calculate $\Omega $ (see \eq{GLA1})   is to find $N\Lb z^+ = z, Y'=Y\Rb$, since
\beq \label{AM3}
\Omega/2\,\,=\,\ \int d^2 x_{12}\,\tau_{pr}\,N\Lb z^+ = z, Y'=Y\Rb
\eeq
However, in order to use \eq{AM3} one needs to know the value of $N$ near to the saturation scale, whereas
our approach has been developed inside of the two regions of saturation scales.
 We use the method related to $t$-channel unitarity (see \fig{calam}) which was  suggested and adjusted to high
density QCD in Refs.\cite{MUSH,IAMU}. In this approach
\beq \label{AM4}
\Delta\Omega\Lb \zeta\Rb/2 \,\,=\,\ \int d z \, \Delta \,N\Lb Y', z\Rb\,\Delta N\Lb Y - Y', z\Rb\,
\eeq
It should be stressed that the ``net diagrams'' of \fig{netset} possess the following remarkable property, that by cutting one BFKL Pomeron
 line we do not change the integration over the kinematic variables that describe other Pomerons.
Therefore, these Pomerons are included  in the description given by equation \eq{AM201} and \eq{AM202} for $N$ and $N^+$. In \fig{calam1} we give an example of the diagram which is described in terms of \eq{AM4}.
This figure illustrates the fact that the topology of the net diagrams (see \fig{netset}) is very essential to the derivation of \eq{AM4}. The Pomeron loops cannot be taken into account using \eq{AM4}.
The net diagrams have another remarkable feature, namely that all paths in the diagram that start from one nucleus, also finish at the other nucleus. In other words, there are no loops in the diagram.
Each path can be cut at some value of rapidity, and contributes to \eq{AM4}.
Keeping only the factor in the exponent we obtain that

\bea \label{AM5}
\Delta\Omega\Lb \zeta\Rb/2 \,\, &=& \,\ \int d
\bar{z} \exp \Big(  - {\cal Z} - {\cal Z}^+\Big)\,\ \\
& = &  \int d \bar{z} \exp \Big(-\,\sqrt{\frac{\bas^2\ln\Lb(Y - Y')/\sqrt{\h\zeta - \bar{z}}\Rb}{\h\zeta - \bar{z}}\,\Lb Y - Y'\Rb^2} \,-\,\sqrt{\frac{\bas^2\ln\Lb Y' /\sqrt{\bar{z}}\Rb}{\bar{z}}\,\Lb Y' \Rb^2}\;\;\;\Big)\nn
\eea
where $ \bar{z}\,=\,z^+ - \h \la_s\,Y'$. Integrating over $\bar{z}$ using the steepest decent method we obtain that the saddle point in the $\bar{z}$-integration is $\bar{z}_{SP} = \zeta/4$, which leads to
\beq \label{AM6}
\Delta\Omega\Lb \zeta\Rb/2 \,\,\,=\,\,\,\tilde{H}\,\exp\Big\{\,-\,\sqrt{\bas^2\ln\Lb \frac{ Y /2}{\sqrt{\zeta/4}}\Rb}\frac{ Y}{\sqrt{\zeta/4}}\;\Big\}
\eeq

It is interesting to note that the dependence on $Y'$  disappears, and the approach to the unitarity bound is much milder than in the case of the solution to the BK equation. Indeed, in the saturation region with two saturation scales $\Delta \Omega \,\propto\, \exp\Big(-\,Const\, \sqrt{Y}\Big)$ while for the BK equation $\Delta \Omega\, \propto \,\exp\Big(-\,Const\, Y^2\Big)$.
The function $\tilde{H}$ absorbs all the pre-exponential factors that depend on $Y$ and impact parameters.
The nucleus-nucleus scattering amplitude can be written using \eq{GLA1} - \eq{GLA3} in the form
\beq \label{AM7}
\mbox{Im} A\Lb Y; B \Rb \,\,=\,\,1 - \exp \Big( -\,\,\int d^2 b \,d^2 b' \Lb \tilde{S}_{A_1}\Lb b\Rb\,\tilde{S}_{A_2}\Lb \vec{B} - \vec{b}\Rb \,-\,\Delta \Omega \Rb\Big)
\eeq
where $b$ is the impact parameter of the nucleon inside  of one nucleus and $B$ is the distance between the centers of two nucleons.
 $b'$ in \eq{AM7} is the impact parameter of the dipole inside of the nucleon. As we have discussed above,
 in our approach we integrated over this impact parameter.  The typical value of $b'$ is our new dimensional parameter,
 that determines the range of energy which we can reach within this approach. Generally speaking the average $b'$ increases
 with energy, but we assume that $<  b' >\,\, \ll\,\, R_A$.
This inequality determines the range of energies where we can trust our approach. The energy dependence of $< b>$ cannot be found in
 our approach, since in the framework of perturbative QCD, $ < b > \propto s^\Delta$. However we know
(see the Froissart theorem in ref. \cite{FROI})
that $< b' >$ can increase only logarithmically, but the parameters of such an increase will depend crucially on the non-perturbative parameter: the mass of the lightest hadron.
The best estimate is to insert into \eq{AM7} the experimental value of the
inelastic cross section for the proton-proton interaction. In doing so we obtain
\beq \label{AM8}
\mbox{Im} A\Lb Y; B \Rb \,\,=\,\,1 - \exp \Big( \,-\,\sigma_{in}\Lb p p \Rb\,\int d^2 b  \tilde{S}_{A_1}\Lb b\Rb\,\tilde{S}_{A_2}\Lb \vec{B} - \vec{b}\Rb\Big)
\eeq
It should be stressed that \eq{AM8} flows naturally from our approach, whereas however it is certainly incorrect in the usual
 Glauber-Gribov \cite{GRIBA} approximation (see Refs.\cite{GG}). However, we would like to stress that the form factors
 $\tilde{S}_{A_i}$ are different from the usual nucleus form factor $S_A\Lb b\Rb$. Indeed,
 $S_A\Lb b\Rb$ gives the number of nucleons that have impact parameter $b$ and it is equal to
\beq \label{AM9}
S_A\Lb b\Rb\,=\,\int^{\infty}_{-\infty} d z \,\rho_A\Lb z,b\Rb
\eeq
where $\rho$ is the density of nucleons in the nucleus.  $\tilde{S}_A\Lb b\Rb$ is equal to
\beq \label{AM10}
\tilde{S}_A\Lb b\Rb\,=\,\int^{R_N}_{-R_N} d z \,\rho_A\Lb z,b\Rb
\eeq
and it  characterizes the situation when all nucleons with the impact parameter $b$ interact as one.

\section{Conclusions}
In this paper  we consider nucleus-nucleus scattering at high energies in the framework of the BFKL
 Pomeron Calculus, given by \eq{BFKLFI}, which follows from the direct sum over Feynman diagrams. It turns out that for the dilute-dense
 system scattering, this approach
gives the same results as other approaches, such as the dipole approach and  the JIMWLK equation. 
 Therefore, it seems reasonable to discuss nucleus scattering  using the BFKL Pomeron Calculus, since other approaches have failed to
successfully  derive the set of equations for the dense-dense scattering amplitude. 
However, we would like to mention that in spite of the fact that we believe that the BFKL Pomeron Calculus describes the high energy
 interaction in QCD, we cannot exclude the opposite.  In particular, it is possible that the four- Pomeron
 interaction could contribute at high energies. The most important result of this paper is the statement that
 at high energies, the equations in the BFKL Pomeron Calculus do not contradict either  the $s$-channel unitarity, or the property of
 crossing symmetry. The main properties of our solution  for the nucleus-nucleus scattering amplitude,
 can be summarized as follows.

\begin{enumerate}
 \item
The contribution of  short distances to the opacity $\Omega$,  dies at high energies.
\item
 The opacity tends to unity at high energies.
\item
The main
contribution that survives, originates from soft (long distance) processes for large values of the impact parameter.
\end{enumerate}

The corrections to the opacity $ \Omega = 1 - \Delta N$  that stem from short distances,
have been discussed in this paper and it was shown that they behave differently from the corrections to the Balitsky-Kovchegov equation.
 Indeed, it turns out that  $\Delta N \,\propto\, \exp\Lb - Const \sqrt{Y}\Rb$ while for the BK equation, $ \Delta N \,\propto\, \exp\Lb - Const Y^2 \Rb$.

The most salient result of this paper is the formula of \eq{AM8} that describes  nucleus-nucleus collisions. This formula is instructive, especially since in the usual Glauber-Gribov approach, there is no reason to expect that this formula works \cite{GG}.

All of these above mentioned results are based on the BFKL Pomeron Calculus . It should be stressed that  this Calculus 
and the action of  \eq{BFKLFI} has been proven only at $N_c \,\gg\,1$.
However, we should emphasize that the equivalence  of this  approach with  other approaches (see refs. \cite{KOLU,HIMST} for example)
that originate from the JIMWLK equation \cite{JIMWLK}, have not been proven.
It should be stressed that the two strategies reflect two different fundamental features of QCD, namely that
the BFKL Pomeron Calculus satisfies, by construction, the $t$-channel unitarity whereas JIMWLK-based approaches
are precise from the point of view of $s$-channel unitarity.
In general \eq{EQ21} and \eq{EQ22} give the equations for the dense-dense system of scattering, in the mean field approximation
 which replaces the BK equation, while the solution that was found in this paper has the same legacy as the solution to the BK equation
 deeply in the saturation region, given in Ref. \cite{LTSOL}.

 However,  the mean field approximation cannot be 
trusted at extremely high energies.  
Strictly speaking starting from $Y = Y_m$ where $\as^2 \exp\Big(\omega(0) Y_m\Big) = 1$, both solutions are not valid and we need to take into account all kind of enhanced diagrams (see \fig{eqAAil}-b for example). However, to calculate such diagrams  we need to find the Green function of the Pomeron in the mean field approximation. Since at $Y = Y_m$  these solutions give the amplitude which is close to the unitarity boundary , the Green function can be small\footnote{The Green function in the mean field approximation has not been found in the BFKL Pomeron calculus. However  numerical estimates \cite{BRAUNGF} and examples of analytical solution for this Green function in the BFKL Pomeron calculus in zero transverse dimension\cite{}GKLM} show that such a scenario looks very plausible.
Therefore,
the problem of taking into account of Pomeron loops  could be not important for the understanding of the
phenomena of 
saturation.

\section*{Acknowledgements}

This work was supported in part by the  Fondecyt (Chile) grant  \# 1100648.
This research was supported by CENTRA, and the Instituto Superior T\'ecnico (IST), Lisbon.
One of us (JM) would like to thank Tel Aviv University for their hospitality on this visit,
during the time of the writing of this paper.


\end{document}